\documentclass[reprint,superscriptaddress,showkeys,showpacs,amsmath,amssymb,aps,prb]{revtex4-2}

\usepackage{graphicx}
\usepackage{dcolumn}
\usepackage{bm}
\usepackage{color}
\usepackage{gensymb}
\definecolor{darkblue}{rgb}{0,0,.6}
\usepackage[colorlinks=true, citecolor=darkblue, linkcolor=darkblue, menucolor=darkblue, urlcolor=darkblue]{hyperref}

\newcommand{\RuCl}{$\alpha$-RuCl$_3$}
\newcommand{\LiIrO}{$\alpha$-Li$_2$IrO$_3$}
\newcommand{\NaIrO}{Na$_2$IrO$_3$}

\begin{document}

%\preprint{arxiv:blah.blah}

\title{Non-local features of the spin-orbit exciton in Kitaev materials}

\author{Blair~W.~Lebert}
\author{Subin~Kim}
\affiliation{Department of Physics, University of Toronto, Toronto, Ontario, M5S 1A7, Canada}

\author{Beom Hyun Kim}
\affiliation{Center for Theoretical Physics of Complex Systems, Institute for Basic Science, Daejeon 34126, Republic of Korea}
\affiliation{Korea Institute for Advanced Study, Seoul 02455, Republic of Korea}

\author{Sae Hwan Chun}
\affiliation{Pohang Accelerator Laboratory, Pohang, Gyeongbuk 37673, Republic of Korea}

\author{Diego Casa}
\affiliation{Advanced Photon Source, Argonne National Laboratory, Argonne, Illinois 60439, USA}

\author{Jaewon Choi}
\author{Stefano Agrestini}
\author{Kejin Zhou}
\author{Mirian Garcia-Fernandez}
\affiliation{Diamond Light Source, Harwell Campus, Didcot OX11 0DE, United Kingdom}

\author{Young-June~Kim}
\affiliation{Department of Physics, University of Toronto, Toronto, Ontario, M5S 1A7, Canada}
\email{youngjune.kim@utoronto.ca}

\date{\today}

\begin{abstract}
A comparative resonant inelastic x-ray scattering (RIXS) study of three well-known Kitaev materials is presented: \LiIrO{}, \NaIrO{}, and \RuCl{}. Despite similar low-energy physics, these materials show distict electronic properties, such as the large difference in the size of the charge gap. The RIXS spectra of the spin-orbit exciton for these materials show remarkably similar three-peak features, including sharp low energy peak (peak A) as well as transitions between $j_{\text{eff}}=1/2$ and $j_{\text{eff}}=3/2$ states. Comparison of experimental spectra with cluster calculations reveals that the observed three-peak structure reflects the significant role that non-local physics plays in the electronic structure of these materials. In particular, the low-energy peak A arises from a holon-doublon pair rather than a conventional particle-hole exciton as proposed earlier. Our study suggests that while spin-orbit assisted Mott insulator is still the best description for these materials, electron itinerancy cannot be ignored when formulating low-energy Hamiltonian of these materials.
\end{abstract}

%\keywords{}
%\pacs{}
\maketitle

\section{Introduction}

Honeycomb Kitaev materials have been studied extensively in search of a Kitaev quantum spin liquid \cite{kitaevAnyonsExactlySolved2006,jackeliMottInsulatorsStrong2009,rauSpinOrbitPhysicsGiving2016,winterModelsMaterialsGeneralized2017a,hermannsPhysicsKitaevModel2018,takagiConceptRealizationKitaev2019,motomeHuntingMajoranaFermions2020a,trebstKitaevMaterials2022,kimRuClOtherKitaev2022}. The main Kitaev candidates are
\LiIrO{} \cite{chaloupkaKitaevHeisenbergModelHoneycomb2010,singhRelevanceHeisenbergKitaevModel2012a},
\NaIrO{} \cite{shitadeQuantumSpinHall2009a, chaloupkaKitaevHeisenbergModelHoneycomb2010,singhAntiferromagneticMottInsulating2010b,singhRelevanceHeisenbergKitaevModel2012a,chaloupkaZigzagMagneticOrder2013,reutherFinitetemperaturePhaseDiagram2011,mazinNaIrOMolecular2012}, and
\RuCl{} \cite{plumbRuClSpinorbitAssisted2014,searsMagneticOrderRuCl2015,banerjeeProximateKitaevQuantum2016a,doMajoranaFermionsKitaev2017,sandilandsScatteringContinuumPossible2015,kimRuClOtherKitaev2022} which all have honeycomb planes formed from edge-sharing IrO$_6$/RuCl$_6$ octahedra \cite{singhAntiferromagneticMottInsulating2010b,omalleyStructurePropertiesOrdered2008,stroganov1957,banerjeeProximateKitaevQuantum2016a,johnsonMonoclinicCrystalStructure2015a,caoLowtemperatureCrystalMagnetic2016}.
Due to the large spin-orbit coupling (SOC), $\lambda$, {the magnetism of these materials is described by $j_{\text{eff}}=1/2$ pseudospins which experience a bond-dependent Kitaev interaction ($K$). In addition, off-diagonal symmetric exchange (so-called $\Gamma$) and isotropic Heisenberg interaction ($J_1$) are necessary to describe these Kitaev materials \cite{Rau2014,Katukuri2014,Chaloupka2015,winterModelsMaterialsGeneralized2017a}. However, it turns out that the $J_1-K-\Gamma$ model is insufficient to explain the observed magnetic ground states, and at least one more interaction term is required. Two leading candidates are the third neighbor Heisenberg interaction $J_3$ and the extra off-diagonal term due to trigonal distortion, $\Gamma^\prime$ \cite{bhattacharjeeSpinOrbitalLocking2012a,Winter2016,Baez2019,Kimchi2015,Lee2016,Gordon2019,searsFerromagneticKitaevInteraction2020a}. While these two interactions can give rise to the zigzag ground state found in \RuCl{} and \NaIrO{}, they represent very different viewpoints on these materials. Unlike $\Gamma^\prime$, which arises from the trigonal distortion in the strong-coupling limit, further neighbor interactions require longer-range hopping and emphasizes the more itinerant nature of electrons \cite{Winter2016}.

The nature of electron itinerancy and the strength of electron correlation in these systems has been debated in earlier studies. While the $j_{\text{eff}}=1/2$ picture arises from the strong correlation limit  \cite{jackeliMottInsulatorsStrong2009}, quasi-molecular orbital (QMO) theory was proposed to describe \NaIrO{} in the weak correlation limit \cite{mazinNaIrOMolecular2012,Foyevtsova2013}.} Experimental studies, in particular, resonant inelastic x-ray scattering (RIXS), have been particularly useful in this debate in favor of the strong correlation.
In their Ir L$_3$-edge RIXS study of \NaIrO{}, \citet{gretarssonCrystalFieldSplittingCorrelation2013} showed a three-peak structure (labeled A, B and C from low to high energy), where peaks B and C are due to intrasite transitions from the $j_{\text{eff}} = 1/2$ to the $j_{\text{eff}} = 3/2$ states. 
%The latter states are split due to small trigonal distortions ($\Delta_t$). 
The fact that these excitations are found near $3\lambda/2$, as expected from atomic calculations and that they show flat dispersion, combined with small splitting between peaks B and C (due to trigonal distortion), supported the spin-orbit (SO) Mott insulator picture.
Meanwhile, peak A at lower energy (just below the charge gap) was attributed to a conventional exciton due to Coulomb interaction between electron and hole ($e$-$h$) \cite{gretarssonCrystalFieldSplittingCorrelation2013,Hermann2019,Comin2012}. This explanation is unsatisfactory, since it resorts to a weak-correlation picture for A while B+C is explained in the strong-correlation picture. An alternate explanation was put forward in the theoretical work by \citet{BHKim2016}. In their numerical study of a three-band Hubbard model, they showed that peak A can be present without explicit $e$-$h$ Coulomb interaction \cite{kimElectronicExcitationsEdgeshared2014b, igarashiCollectiveExcitationsNa2016a, igarashiResonantInelasticXray2016a, BHKim2016}.

In this paper, we revisit the RIXS spectra of spin-orbit excitations in honeycomb Kitaev materials to resolve the debate about electron correlation in these materials. Specifically, we carried out new RIXS measurements on \RuCl{} and \LiIrO{} single crystals using higher energy resolution of 35~meV. We note that the charge gap in \RuCl{} is well separated from the spin-orbit energy scale and \LiIrO{} has the magnetic ground state distinct from the other two, thus each material brings properties distinct from the well-studied \NaIrO{}. We find that the RIXS spectra of both samples are remarkably similar to the three-peak structure previously observed for \NaIrO{} \cite{gretarssonCrystalFieldSplittingCorrelation2013}, indicating that the full three-peak structure comprises the spin-orbit exciton spectral feature. To investigate the origin of peak A, we performed cluster calculations on three-band Hubbard model. While the experimental three-peak feature is reproduced well with the calculation, the absence of peak A in our single-site calculation indicates intersite nature of peak A. Our result therefore reveals the importance of hopping, or ``non-local'' physics in these materials, {which suggest importance of longer-range interactions, such as $J_3$, in the magnetic Hamiltonian.}

\begin{figure}[t]
\includegraphics{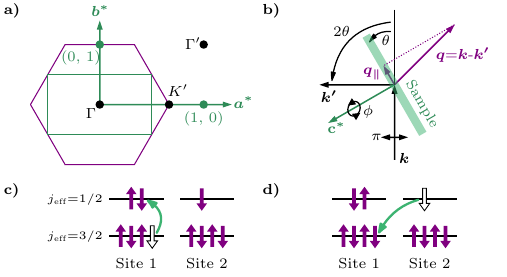}
\caption{\label{fig:overview}
a) In-plane reciprocal space showing hexagonal Brillouin zone (purple) and projection of $C2/m$ Brillouin zone (green). The $C2/m$ lattice is used throughout this paper and the projection of its $\bm{a^\ast} = (h, 0)$ and $\bm{b^\ast} = (0,k)$ axes are shown as green vectors. b) RIXS scattering plane used for experiments and calculations. Incoming x-rays with $\pi$ polarization and momentum $\bm{k}$ are scattered at $2\theta=90\degree$ with momentum $\bm{k^\prime}$ transferring $\bm{q}=\bm{k}-\bm{k^\prime}$ momentum to the sample (green bar). The sample is mounted with the honeycomb plane at an angle $\theta$ leading to transferred in-plane momentum $\bm{q_\parallel}$. The azimuthal angle $\phi$ rotates the sample around the $\bm{c^\ast}$ axis, with $\bm{a^\ast}$ ($\bm{b^\ast}$) in the scattering plane at $\phi=0\degree$ ($\phi=90\degree$). c) \& d) Schematic energy diagram of two excitation processes. Filled and unfilled arrows represent electrons and holes respectively, where up (down) arrows denote Kramer's doublet with positive (negative) eigenvalues of $j^z_{\text{eff}}$. See text for details.}
\end{figure}

\section{Methods}

High-quality \RuCl{} \cite{plumbRuClSpinorbitAssisted2014,searsMagneticOrderRuCl2015} and \LiIrO{} \cite{linVaporGrowthChemical2012,freundSingleCrystalGrowth2016} single crystals were grown by the vapor transport method and have previously been characterized \cite{lebertAcousticPhononDispersion2022,chunOpticalMagnonsDominant2021a}. Single domains of \LiIrO{} were found using magnetic Bragg peaks as described in Ref.~\cite{chunOpticalMagnonsDominant2021a}. In this paper, we use $C2/m$ notation and reciprocal lattice units to describe the honeycomb materials (Fig.~\ref{fig:overview}a). We use our previously published data for \NaIrO{} \cite{gretarssonCrystalFieldSplittingCorrelation2013}.

All RIXS measurements were taken in horizontal scattering geometry with $2\theta=90^{\circ}$ and incident $\pi$ polarization to minimize elastic scattering (Fig.~\ref{fig:overview}b). Only the in-plane momentum component shown as $\bm{q_\parallel} = (h,k)$ is considered due to the quasi-2D nature of these materials \cite{gretarssonCrystalFieldSplittingCorrelation2013}. The $\bm{a^\ast}$ axis has an out-of-plane component, but this does not influence our measurements \cite{gretarssonCrystalFieldSplittingCorrelation2013}. Likewise, the RIXS data shows very little temperature dependence (see Appendix A), and only low-temperature data are discussed in the main text. \cite{williamsIncommensurateCounterrotatingMagnetic2016a, singhRelevanceHeisenbergKitaevModel2012a, liuLongrangeMagneticOrdering2011a, mehlawatHeatCapacityEvidence2017, yeDirectEvidenceZigzag2012a, searsPhaseDiagramRuCl2017a, kimRuClOtherKitaev2022}

The Ir L$_3$-edge RIXS (11.215~keV) experiment on \LiIrO{} was performed on the 27-ID-B endstation (MERIX) at the Advanced Photon Source of Argonne National Laboratory \cite{shvydkoMERIXNextGeneration2013}. Incident photons were monochromatized with a Si(844) channel-cut monochromator and scattered photons were analyzed using a spherically bent diced Si(844) analyzer on a 2-meter Rowland circle giving 35~meV FWHM total energy resolution. Ru M$_3$-edge RIXS (459.5~eV) was performed on the I21 beamline at the Diamond Light Source \cite{zhouI21AdvancedHighresolution2022}. Incident photons were monochromatized with a 1000~$\ell$/mm variable line spacing plane grating and 50~$\mu$m exit slit, while scattered photons were analyzed using a 1500~$\ell$/mm spherical variable line spacing grating and a CCD with $13.5 \times 13.5~\mu$m$^2$ pixels at $30 \degree$ grazing angle, resulting in 35~meV FWHM total energy resolution.

Calculations were performed using the three-band ($t_{2g}$-band) Hubbard model of clusters incorporating the electronic hopping among nearest neighbor $t_{2g}$ orbitals, spin-orbit coupling (SOC), trigonal distortion, and Kanamori-type Coulomb interactions (see the details in Refs.~[\onlinecite{BHKim2016}] and~[\onlinecite{Lee2021}]). The 1-site and 4-site calculations used an open boundary condition, while the 6-site calculation used a periodic boundary condition. To calculate the RIXS spectra, we employed the Kramers-Heisenberg formula with the fast collision approximation and dipole approximation \cite{Ament2011}. The RIXS spectra are calculated with the same geometry as each individual measurement, i.e. the same $2\theta$, $\theta$, and $\phi$ as described in Fig.~\ref{fig:overview}b. To match the experiments, RIXS spectra are calculated with $\pi$ incoming polarization (defined as the polarization vector in the scattering plane) and with a 1:1 ratio of $\pi$:$\sigma$ for the outgoing polarization since it is not analyzed in our experimental setup. Finally, we convolve the RIXS calculations with a 35~meV resolution function. Physical parameters used in our calculations are given further below and more details can be found in Appendix B.

\begin{figure}[t]
\includegraphics{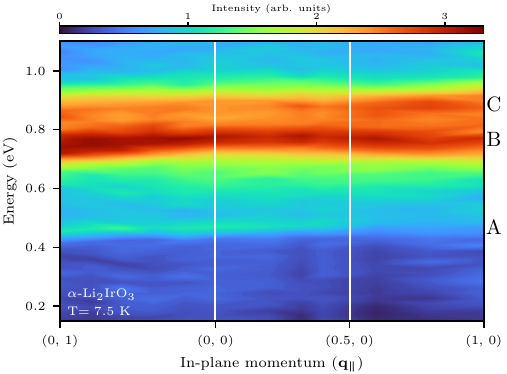}
\caption{\label{fig:Li_cmap}
In-plane momentum dependence of excitations in \LiIrO{} at 7.5~K measured with Ir L$_3$-edge RIXS. The main peaks are labelled from low to high energy as A, B, and C.
}
\end{figure}

\section{Results and Discussion}

Ir L$_3$-edge RIXS spectra of \LiIrO{} are shown in Fig.~\ref{fig:Li_cmap} along two high-symmetry directions measured at 7.5~K. This result is very similar to the \NaIrO{} spectra reported in Ref.~\cite{gretarssonCrystalFieldSplittingCorrelation2013}, described by the characteristic three peak structure on top of a background of continuum excitations. All peaks show small dispersion of about 20~meV bandwidths. Although it is tempting to attribute the observed dispersive behavior of these peaks to magnetic energy scale, further systematic study with higher resolution will be required to investigate the dispersion, and will not be discussed further in the current paper. The peak positions found in the current single crystal study is slightly different from the powder data in Ref.~\cite{gretarssonCrystalFieldSplittingCorrelation2013}, likely due to powder averaging. Overall the peaks in \LiIrO{} are 7--11\% higher in energy with respect to \NaIrO{}. This result also agrees well with the O K-edge RIXS data reported in \cite{valeHighresolutionResonantInelastic2019b} as summarized in Table~\ref{fits}.

\begin{table}[t]
\caption{Measured excitation energies at $(0,0)$ given in eV. Unless noted with (p), all samples are single crystals.}
\label{fits}
\begin{ruledtabular}
\begin{tabular}{l c c c c}
& Technique & Peak A & Peak B & Peak C \\
\hline
$\alpha$-Li$_2$IrO$_3$ & Ir L$_3$ RIXS & 0.47(1) & 0.77(1) & 0.89(1) \\
$\alpha$-Li$_2$IrO$_3$ (p) \cite{gretarssonCrystalFieldSplittingCorrelation2013} & Ir L$_3$ RIXS & 0.45(2) & 0.72(2) & 0.83(2) \\
$\alpha$-Li$_2$IrO$_3$ \cite{valeHighresolutionResonantInelastic2019b} & O K RIXS & 0.46(2) & 0.77(2) & 0.88(2) \\
\hline
Na$_2$IrO$_3$ \cite{gretarssonCrystalFieldSplittingCorrelation2013} & Ir L$_3$ RIXS & 0.42(1) & 0.72(2) & 0.83(2) \\
\hline
$\alpha$-RuCl$_3$ & Ru M$_3$ RIXS & 0.159(2) & 0.245(6) & 0.289(9) \\
$\alpha$-RuCl$_3$ \cite{warzanowskiMultipleSpinorbitExcitons2020b} & Raman & & 0.248(1) & 0.290(4) \\
$\alpha$-RuCl$_3$ \cite{suzukiProximateFerromagneticState2021a} & Ru L$_3$ RIXS &  & 0.25(2) & \\
$\alpha$-RuCl$_3$ \cite{lebertResonantInelasticXray2020} & Ru M$_3$ RIXS &  & 0.231(3) &
\end{tabular}
\end{ruledtabular}
\end{table}

\begin{figure*}[t]
\includegraphics{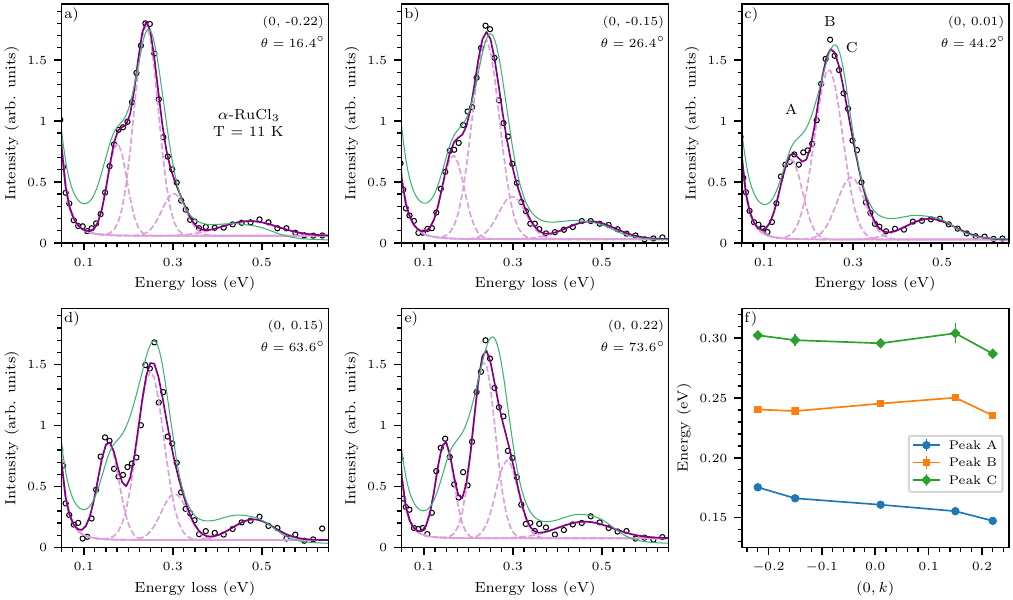}
\caption{\label{fig:RuCl}
Experimental Ru M$_3$-edge RIXS data and fits of \RuCl{} at 11~K compared to calculations. Spectra were taken with $\bm{b^\ast}$ axis in scattering plane ($\phi=90\degree$) while varying $\theta$ from grazing incidence (panel a) to grazing emission (panel e) with corresponding $\bm{q_\parallel}$ and $\theta$ shown in the top right corner of each panel. Data are shown as black circles and fits are shown as solid purple lines, with fit components as dashed purple lines: in particular peaks A, B, and C which are labelled in c), as well as an additional higher energy excitation. RIXS calculations are shown as green lines for $\Delta_t < 0$. The momentum points and scattering geometry are decoupled in our calculations, therefore since all Ru M$_3$-edge measurements are close to $(0,0)$ we calculate the RIXS spectra at $(0,0)$ while using the experimental $\theta$ angle. {The momentum dependence of the extracted peak positions are plotted in f).}
}
\end{figure*}

In Fig.~\ref{fig:RuCl}, we show Ru M$_3$-edge RIXS spectra which clearly confirm the existence of peak A in \RuCl{}. The splitting between the B and C peaks is not resolved with our resolution, however a shoulder feature is visible, for example, in the (0, 0.22) spectrum. Therefore, we fitted our results with the small B-C splitting. The peak assignment is supported by the good agreement between our extracted energies at $(0,0)$  and the Raman results (see Table~\ref{fits}).
In their previous Ru L$_3$-edge RIXS study, \citet{suzukiProximateFerromagneticState2021a} observed a single peak around $\approx$250~meV (Table~\ref{fits}), but were unable to resolve any splitting due to $\Delta_t$ because of the coarse (100~meV) resolution. They did however note excessive spectral weight at lower energy which they associated with peak A.

The existence and possible origin of peak A however remains a mystery. The Mott gap in \RuCl{} is widely debated with values ranging from 0.2-2.2~eV \cite{guizzettiFundamentalOpticalProperties1979,rojasHallEffectARuCl31983,polliniPhotoemissionStudyElectronic1994,plumbRuClSpinorbitAssisted2014,kimKitaevMagnetismHoneycomb2015a,Sandilands2016,sandilandsSpinorbitExcitationsElectronic2016a,koitzschEffDescriptionHoneycomb2016a,Sinn2016,ziatdinovAtomicscaleObservationStructural2016a,zhouAngleresolvedPhotoemissionStudy2016b,reschkeElectronicPhononExcitations2017a,biesnerDetuningHoneycombRuCl2018,warzanowskiMultipleSpinorbitExcitons2020b,koitzschLowtemperatureEnhancementFerromagnetic2020,reschkeSubgapOpticalResponse2018a,nevolaTimescalesExcitedState2021}, however the consensus seems to be 0.9--1.2~eV \cite{guizzettiFundamentalOpticalProperties1979,polliniPhotoemissionStudyElectronic1994,kimKitaevMagnetismHoneycomb2015a, Sandilands2016, sandilandsSpinorbitExcitationsElectronic2016a,koitzschEffDescriptionHoneycomb2016a,zhouAngleresolvedPhotoemissionStudy2016b,biesnerDetuningHoneycombRuCl2018,warzanowskiMultipleSpinorbitExcitons2020b,koitzschLowtemperatureEnhancementFerromagnetic2020}. This implies that if peak A exists in the RIXS spectra of \RuCl{}, then it cannot be an $e$-$h$ excitation across the Mott gap. In fact, \citet{BHKim2016} explicitly predict that peak A exists in \RuCl{} and is unrelated to the Mott gap. They propose that peak A is a doublon-holon excitation created by a $j_{\text{eff}} = 1/2 \rightarrow j_{\text{eff}} = 3/2$ intersite process (Fig.~\ref{fig:overview}d) coupling with the RIXS active $j_{\text{eff}} = 3/2 \rightarrow j_{\text{eff}} = 1/2$ intrasite process (Fig.~\ref{fig:overview}c).

The small momentum transfer of Ru M$_3$-edge RIXS limits our exploration of reciprocal space \cite{lebertResonantInelasticXray2020}. Still, we see some hints of dispersion especially towards grazing emission geometry, seen in Fig.~\ref{fig:RuCl}e), where peak A is most clearly resolved. The peak positions are plotted as a function of momentum in Fig.~\ref{fig:RuCl}f). However, we would like to point out that the matrix-element effect might be partially responsible for the observed momentum dependence. To illustrate this point, we plot our calculations (see below) as green lines in Fig.~\ref{fig:RuCl}, where all the calculations are performed at $(0,0)$ but independently varying the RIXS geometry to match experiment. 
The calculations capture qualitatively the redistribution of spectral weight towards lower energies going from grazing incidence to grazing emission.
{However, the agreement becomes progressively worse as $\theta$ increases, with peak A shifting its position by almost 30 meV. Further studies, such as outgoing photon polarization analysis and calculations using larger clusters will be necessary for quantitative understanding of this observation.} 

\begin{figure}[t]
\includegraphics{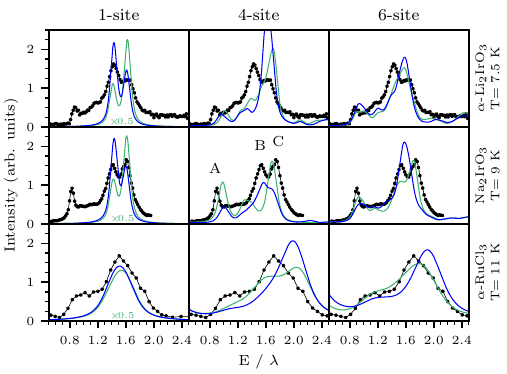}
\caption{\label{fig:calcs} Comparison between experimental (black) and calculated RIXS spectra at $(0,0)$ with $\Delta_t > 0$ (blue) and $\Delta_t < 0$ (green). The rows correspond to different materials and the columns to calculations with different number of sites. The energy axis has been scaled by the spin-orbit coupling ($\lambda$) used in the calculation for each compound (Table~\ref{para}). The 1-site calculated spectra's intensities have been scaled by half for visibility. The A, B, and C peaks are labelled in the middle panel.
}
\end{figure}

To illustrate the similarity of RIXS features in three Kitaev materials, we compare experimental and calculated RIXS spectra for \LiIrO{}, \NaIrO{}, and \RuCl{} at $(0,0)$ in Fig.~\ref{fig:calcs}. The energy axes have been scaled by $\lambda$ used in our calculations and we observe robust scaling behavior indicating very similar SO Mott insulator physics.

\begin{table}[b]
\caption{Physical parameters for the spin-orbit couplings of the 1-site and 4/6-site calculations $\lambda_0$ and $\lambda$, trigonal distortion $\Delta_{t}$ ($E_{a_{1g}}-E_{e_g'}$), Coulomb repulsion $U$, and hopping integrals $t_1$ to $t_4$ used in 4/6-site calculations given in meV. $J_H=350$~meV for all compounds. }
\label{para}
\begin{ruledtabular}
\begin{tabular}{l c c c c c c c c}
& $\lambda_0$ & $\lambda$ & $\Delta_{t}$ & $U$ & $t_1$ & $t_2$ & $t_3$ & $t_4$ \\
\hline
$\alpha$-Li$_2$IrO$_3$ & $540$ & $540$ & $\pm160$ & 1800 & $72.0$ & $251.2$ & $-136.1$ & $-49.6$ \\
Na$_2$IrO$_3$ & $510$ & $470$ & $\pm180$ & 2000 & $44.3$ & $321.1$ & $-4.9$ & $-23.4$ \\
$\alpha$-RuCl$_3$ & $165$ & $145$ & $\pm40$ & 2350 & $54.3$ & $185.0$ & $-138.1$ & $-17.7$ \\
\end{tabular}
\end{ruledtabular}
\end{table}

The physical parameters used in the calculations are shown in Table~\ref{para}. The $\lambda_0$ and $\Delta_{t}$ values were chosen by matching the energy positions of peaks B and C in 1-site calculations to the experimental peaks (left column of Fig.~\ref{fig:calcs}). Both positive and negative $\Delta_t$ of equal magnitude were calculated and are shown as blue and green lines respectively. Since we are only considering $t_{2g}$ orbitals, the $\lambda_0$ is an effective $\lambda$ and we found that slightly different $\lambda_0$ values are found for \NaIrO{} and \LiIrO{}, presumably due to the slight difference in lattice environments. In addition, we ended up adopting slightly different values for $\lambda$ in 4- and 6-site calculations to improve the fit, because non-local effects (described below) changes the positions of peak B and C from $\frac{3}{2}\lambda_0$~\cite{Lee2021}.
$U$ was selected to fit the optical conductivity while keeping Hund's coupling at $J_H = 350$~meV \cite{Sinn2016} (see Appendix C). Hopping parameters were adopted from \citet{Winter2016} with averaging in order to preserve the three-fold rotational symmetry of our cluster. We also scaled the hopping parameters by 104\% for \LiIrO{}, 120\% for \NaIrO{}, and 115\% for \RuCl{} to better match the energy position of peak A.

The 1-site calculations (left column) show only two peaks: B and C. These peaks are created in the direct RIXS process summarized in Fig.~\ref{fig:overview}c, where a core electron is excited to $j_{\text{eff}} = 1/2$ while an electron from $j_{\text{eff}} = 3/2$ falls to fill the core hole. This is effectively a local $j_{\text{eff}} = 3/2$ to $j_{\text{eff}} = 1/2$ transition which leaves behind a hole in $j_{\text{eff}} = 3/2$. Note that the trigonal distortion is omitted in Fig.~\ref{fig:overview}c-d for simplicity. The lack of peak A in single-site calculations indicates that local (intrasite) physics is insufficient to describe the observed spectra and an additional intersite interaction is required. Our attempt to distinguish between positive and negative $\Delta_t$ by comparing calculations with experimental spectra is inconclusive as shown in Fig.~\ref{fig:calcs} with blue and green lines. \citet{chaloupkaMagneticAnisotropyKitaev2016b} present a method of determining the sign of $\Delta_t$ by comparing the intensity of B and C peak (using neutron scattering). When you include intersite hopping, B and C excitations are not easy to separate and the neat distinction between the two using the atomic picture is no longer valid. 

Peak A is found only when intersite hopping is included in 4-site (middle column) and 6-site (right column) calculations. The final state of a possible intersite process is shown in Fig.~\ref{fig:overview}d. This describes a higher-order process, in which the $j_{\text{eff}} = 3/2$ hole left on site 1, after intrasite $j_{\text{eff}} = 3/2 \rightarrow j_{\text{eff}} = 1/2$ process (i.e., Fig.~\ref{fig:overview}c), is filled by $j_{\text{eff}} = 1/2$ electron from site 2. We note that an equivalent intersite $j_{\text{eff}} = 1/2 \rightarrow j_{\text{eff}} = 1/2$ direct hopping process is strongly suppressed in these edge-shared honeycomb materials because of the nearly $90\degree$ Ir-O-Ir or Ru-Cl-Ru bond angle. The final state depicted in Fig.~\ref{fig:overview}d is a doublon-holon pair that could give rise to peak A in RIXS spectra, as suggested in Ref.~\cite{BHKim2016}.
The doublon-holon peak is lower in energy with respect to the intrasite SO excitons due to the kinetic energy gained from hopping \cite{BHKim2016}. We repeated the calculation for a 4-site cluster, which does not support the QMO state due to the absence of ring geometry \cite{mazinNaIrOMolecular2012}, while still allowing intersite excitations. The persistence of peak A even in 4-site calculations, therefore, strongly supports that these compounds can be described as SO Mott insulators, even though not quite in the limit of strong correlation. The importance of hybridization and hopping in the RIXS spectra was also pointed out by \citet{delatorre2021} in their study of Ag$_3$LiIr$_2$O$_6$.

%%%%%%%%%%%%%%%%%% %%%%%%%%%%%%%%%%%% %%%%%%%%%%%%%%%%%% %%%%%%%%%%%%%%%%%% %%%%%%%%%%%%%%%%%%
%%%%%%%%%%%%%%%%%% %%%%%%%%%%%%%%%%%% %%%%%%%%%%%%%%%%%% %%%%%%%%%%%%%%%%%% %%%%%%%%%%%%%%%%%%

\section{Conclusions}
We have performed RIXS on single crystals of \RuCl{} and \LiIrO{} which complement earlier data on \NaIrO{} \cite{gretarssonCrystalFieldSplittingCorrelation2013}. The discovery of peak A in \RuCl{} shows the ubiquitousness of this excitation in honeycomb Kitaev materials, whether they be $4d$ or $5d$, even when the Mott gap is not involved.
Our calculations show the importance of intersite hopping in producing peak A, which is attributed to a doublon-holon excitation. Furthermore, our calculations emphasize that all three compounds can be described with intermediate correlation. While electron itinerancy is important, these systems still remain in the realm of spin-orbit Mott insulator, { and the non-local effect can be revealed through the longer range magnetic interaction $J_3$. Our work is a first step --- quantitative 
description of full non-local effects will require calculations including contributions from $e_g$ orbitals and hopping integrals between further neighbors. In addition, detailed measurements of magnon dispersion relation will be desirable for Ir-based Kitaev materials, which is currently limited by RIXS instrumental energy resolution \cite{gretarssonMagneticExcitationSpectrum2013a,kimDynamicSpinCorrelations2020a}.}

\begin{acknowledgments}
We thank J. F. Mitchell of Argonne National Laboratory for helpful discussions on crystal growth of \LiIrO{}. Work at the University of Toronto was supported by the Natural Science and Engineering Research Council (NSERC) of Canada, Canadian Foundation for Innovation, and Ontario Innovation Trust. This research used resources of the Advanced Photon Source (APS), a DOE Office of Science User Facility operated for the DOE Office of Science by Argonne National Laboratory under Contract No. DE-AC02-06CH11357. We acknowledge Diamond Light Source for time on beamline I21-RIXS under Proposal MM25479 and MM29055. B.W.L. acknowledges funding from NSERC [funding reference number PDF-546035-2020]. B.H.K. was supported by the Institute for Basic Science in Korea (IBS-R024-D1) and KIAS Individual Grants (CG068702). Numerical computations have been performed with the Center for Advanced Computation Linux Cluster System at KIAS. S.H.C. acknowledges the support by National Research Foundation of Korea (2019R1C1C1010034 and 2019K1A3A7A09033399).
\end{acknowledgments}

\appendix

\section{Temperature dependence of the RIXS spectra}

\begin{figure}[h]
\centering
\includegraphics[width=.8\columnwidth]{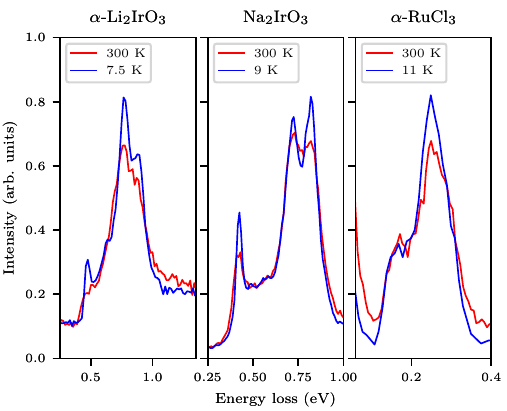}
\caption{Comparison of the RIXS spectra at the room temperature and the base temperature. Note that features are sharper and background is lower at the base temperature, but overall lineshape and the peak positions are unchanged between the two temperatures.
}
\label{fig_temp}
\end{figure}

\section{Theoretical RIXS calculation}

To calculate the theoretical RIXS spectra, we have employed 
the three-band ($t_{2g}$-band) Hubbard model of six-site periodic cluster 
incorporating the electronic hopping among nearest neighboring $t_{2g}$ orbitals,
spin-orbit coupling (SOC), trigonal distortion, and 
Kanamori-type Coulomb interactions 
(see the details in Refs.~[\onlinecite{BHKim2016}] and~[\onlinecite{Lee2021}]).
The RIXS peak of spin-orbit exciton is determined by the SOC parameter $\lambda$.
$\lambda$ is determined to fit the positions of main RIXS peaks well.
Because main peak of optical conductivity in the Mott insulator of 
$t_{2g}$ orbitals are attributed to $U – 3J_H$ parameters~\cite{BHKim2016},
we set $J_H = 0.35$ eV as the estimated value of $\alpha$-RuCl$_3$~\cite{Sinn2016}
and $U$ was selected to fit the optical conductivity well~(see~Fig.~\ref{fig_oc}).
Hopping parameters were adopted in Ref.~[\onlinecite{Winter2016}].
To keep the three-fold rotational symmetry of cluster,
hopping integrals characterized by four parameters $t_1$, $t_2$, $t_3$, and $t_4$
were determined by average values of those in the $C2/m$ structures 
over three neighboring directions.
Because the theoretical RIXS peak A estimated with hopping parameters 
without scale is slightly higher than experimental that,
we enhanced the hopping parameters {up to 104\% for $\alpha$-Li$_2$IrO$_3$}, 120\% for Na$_2$IrO$_3$, and 115\% for $\alpha$-RuCl$_3$, , respectively.
The trigonal distortion parameter $\Delta_{t}$ (=$E_{a_{1g}}-E_{e_g'}$)
was also set to fit the splitting of RIXS peaks of spin-orbit exciton well.
The physical parameters are presented in Table~\ref{para}.

To calculate the RIXS spectra, we employed the Kramers-Heisenberg formula 
with the fast collision approximation and dipole approximation as follow~\cite{Ament2011}:
\begin{widetext}
\begin{equation}
\label{RIXS_eq}
I\left(\omega,\mathbf{q},\bm{\epsilon},\bm{\epsilon^{\prime}} \right)\\
\sim -\frac{1}{\pi} \textrm{Im}\Big[
 \langle \Psi_g| R(\bm{\epsilon}^{\prime},\bm{\epsilon},\mathbf{q})^\dagger
 \frac{1}{\omega-H+E_g+i\delta}
 R(\bm{\epsilon}^{\prime},\bm{\epsilon},\mathbf{q}) |\Psi_g\rangle
\Big],
\end{equation}
\end{widetext}
$\left|\Psi_g\right>$ and $E_g$ are the ground state and its energy, respectively.
$\bm{\epsilon}^{\prime}$ and $\bm{\epsilon}$ are the polarization vectors
of incoming and outgoing x-rays, respectively, and 
the RIXS scattering operator $R(\bm{\epsilon}^{\prime},\bm{\epsilon},\mathbf{q})$ is given as
\begin{equation}
\label{R_eq}
R(\bm{\epsilon}^{\prime},\bm{\epsilon},\mathbf{q})=
\sum_i\sum_{\nu,\nu^{\prime},\sigma} e^{i\mathbf{q}\cdot\mathbf{r}_i}
 T_{\nu^{\prime}\nu}(\bm{\epsilon}^{\prime},\bm{\epsilon})
 c_{i\nu^{\prime}\sigma}c_{i\nu\sigma}^{\dagger},
\end{equation}
where 
   $T_{\nu^{\prime}\nu}(\bm{\epsilon}^{\prime},\bm{\epsilon})=
 \sum_{s} \langle \phi_s| \bm{\epsilon}^{\prime}\cdot \mathbf{r} 
 |\psi_{\nu^{\prime}} \rangle
  \langle \psi_{\nu} | \bm{\epsilon}\cdot \mathbf{r} |\phi_s\rangle$ and
$\mathbf{r}$ is the position operator of valence and core-hole electrons,
${\mathbf r}_i$ is the position vector of lattice site $i$,
$\psi_\nu$ is the local atomic wave function for $t_{2g}$ orbital $\nu$, and 
$\phi_s$ refers to the core-hole wave function ($2p_{3/2}$ for the $L_3$-edge spectrum).

We considered the experimental x-ray geometry
% in which both incoming and outgoing x-rays are 45-degree-away 
% from the normal vector of a honeycomb plane and
% lie in the plane perpendicular to one neighboring bond direction.
%We also assumed the incoming x-ray has the $\pi$ polarization but
and assumed the incoming x-ray has the $\pi$ polarization but
the outgoing x-ray has an arbitrary direction.
The ground state and energy were calculated by the Lanczos method~\cite{Wu2000}.
We performed additional 300 Lanczos iterations with 
an initial state of 
$R(\bm{\epsilon}^{\prime},\bm{\epsilon},\mathbf{q})\left| \Psi_g \right>$.
To keep the orthonormality, we did the Gram-Schmidt orthonornalization for each Lanczos step.
The broadening parameter $\delta$ in Eq.~\ref{RIXS_eq} is set to be 0.02 eV.

\section{Theoretical Optical Conductivity}

\begin{figure*}
\centering
\includegraphics{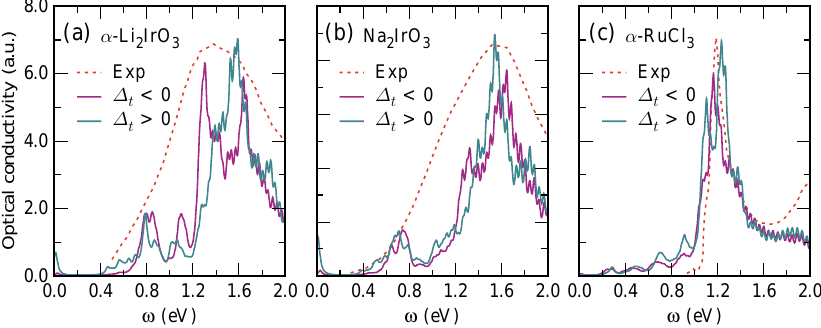}
\caption{ Theoretical and experimental optical conductivity spectra
for (a) $\alpha$-RuCl$_3$, (b) Na$_2$IrO$_3$, and
(c) $\alpha$-Li$_2$RuO$_3$. Solid and dotted lines refer to the
theoretical and experimental spectra, respectively.
Theoretical spectra are calculated with physical prameters in Table~\ref{para},
while experimental those are obtained in Ref.~[\onlinecite{Sandilands2016}] 
for $\alpha$-RuCl$_3$, Ref.~[\onlinecite{Comin2012}] for Na$_2$IrO$_3$,
and Ref.~[\onlinecite{Hermann2019}] for $\alpha$-Li$_2$Ir$_3$, respectively.
}
\label{fig_oc}
\end{figure*}

To verify the relevant physical parameters for $\alpha$-RuCl$_3$,
Na$_2$IrO$_3$, and $\alpha$-Li$_2$IrO$_3$, we also calculated 
the optical conductivity by using the Kubo's formula:
\begin{widetext}
\begin{equation}
\sigma(\omega) \sim \frac{1}{\omega}
  \sum_{n} |\langle \Psi_n|\hat J |\Psi_0\rangle|^2 
  \delta (\omega-E_n\!+\!E_g) 
   = - \frac{1}{\pi}  \ \textrm{Im} 
  \sum_{n} \frac{|\langle \Psi_n|\hat J |\Psi_g\rangle|^2} 
  {(E_n-E_g)(\omega-E_n\!+\!E_g+i\delta_c)},
\label{eqOC}
\end{equation}
\end{widetext}
where $\hat{J}$ is the current operator.
Excited state $\left| \Psi_n \right>$ and
its energy $E_n$ are approximately estimated with the help of
the Lanczos iterations with an initial vector of $ \hat{J} \left| \Psi_g\right>$. 
The broadening parameter $\delta_b$ is set to be 0.02 eV.
As shown in Fig.~\ref{fig_oc},
the main optical peaks at around 1.2 eV for $\alpha$-RuCl$_3$,
1.6 eV for Na$_2$IrO$_3$, and 1.4 eV for $\alpha$-Li$_2$IrO$_3$
are well simulated by the calculation with the physical parameters 
presented in Table~\ref{para}, respectively.

%%%%%%%%%%%%%%%%%%%%%%%%%%%%%%%%%%%
%%%%%%%%%%%%%%%%%%%%%%%%%%%%%%%%%%%

%\bibliography{main}

\begin{thebibliography}{80}%
\makeatletter
\providecommand \@ifxundefined [1]{%
 \@ifx{#1\undefined}
}%
\providecommand \@ifnum [1]{%
 \ifnum #1\expandafter \@firstoftwo
 \else \expandafter \@secondoftwo
 \fi
}%
\providecommand \@ifx [1]{%
 \ifx #1\expandafter \@firstoftwo
 \else \expandafter \@secondoftwo
 \fi
}%
\providecommand \natexlab [1]{#1}%
\providecommand \enquote  [1]{``#1''}%
\providecommand \bibnamefont  [1]{#1}%
\providecommand \bibfnamefont [1]{#1}%
\providecommand \citenamefont [1]{#1}%
\providecommand \href@noop [0]{\@secondoftwo}%
\providecommand \href [0]{\begingroup \@sanitize@url \@href}%
\providecommand \@href[1]{\@@startlink{#1}\@@href}%
\providecommand \@@href[1]{\endgroup#1\@@endlink}%
\providecommand \@sanitize@url [0]{\catcode `\\12\catcode `\$12\catcode
  `\&12\catcode `\#12\catcode `\^12\catcode `\_12\catcode `\%12\relax}%
\providecommand \@@startlink[1]{}%
\providecommand \@@endlink[0]{}%
\providecommand \url  [0]{\begingroup\@sanitize@url \@url }%
\providecommand \@url [1]{\endgroup\@href {#1}{\urlprefix }}%
\providecommand \urlprefix  [0]{URL }%
\providecommand \Eprint [0]{\href }%
\providecommand \doibase [0]{http://dx.doi.org/}%
\providecommand \selectlanguage [0]{\@gobble}%
\providecommand \bibinfo  [0]{\@secondoftwo}%
\providecommand \bibfield  [0]{\@secondoftwo}%
\providecommand \translation [1]{[#1]}%
\providecommand \BibitemOpen [0]{}%
\providecommand \bibitemStop [0]{}%
\providecommand \bibitemNoStop [0]{.\EOS\space}%
\providecommand \EOS [0]{\spacefactor3000\relax}%
\providecommand \BibitemShut  [1]{\csname bibitem#1\endcsname}%
\let\auto@bib@innerbib\@empty
%</preamble>
\bibitem [{\citenamefont {Kitaev}(2006)}]{kitaevAnyonsExactlySolved2006}%
  \BibitemOpen
  \bibfield  {author} {\bibinfo {author} {\bibfnamefont {A.}~\bibnamefont
  {Kitaev}},\ }\href {\doibase 10.1016/j.aop.2005.10.005} {\bibfield  {journal}
  {\bibinfo  {journal} {Annals of Physics}\ }\textbf {\bibinfo {volume}
  {321}},\ \bibinfo {pages} {2} (\bibinfo {year} {2006})}\BibitemShut {NoStop}%
\bibitem [{\citenamefont {Jackeli}\ and\ \citenamefont
  {Khaliullin}(2009)}]{jackeliMottInsulatorsStrong2009}%
  \BibitemOpen
  \bibfield  {author} {\bibinfo {author} {\bibfnamefont {G.}~\bibnamefont
  {Jackeli}}\ and\ \bibinfo {author} {\bibfnamefont {G.}~\bibnamefont
  {Khaliullin}},\ }\href {\doibase 10.1103/PhysRevLett.102.017205} {\bibfield
  {journal} {\bibinfo  {journal} {Phys. Rev. Lett.}\ }\textbf {\bibinfo
  {volume} {102}},\ \bibinfo {pages} {017205} (\bibinfo {year}
  {2009})}\BibitemShut {NoStop}%
\bibitem [{\citenamefont {Rau}\ \emph {et~al.}(2016)\citenamefont {Rau},
  \citenamefont {Lee},\ and\ \citenamefont
  {Kee}}]{rauSpinOrbitPhysicsGiving2016}%
  \BibitemOpen
  \bibfield  {author} {\bibinfo {author} {\bibfnamefont {J.~G.}\ \bibnamefont
  {Rau}}, \bibinfo {author} {\bibfnamefont {E.~K.-H.}\ \bibnamefont {Lee}}, \
  and\ \bibinfo {author} {\bibfnamefont {H.-Y.}\ \bibnamefont {Kee}},\ }\href
  {\doibase 10.1146/annurev-conmatphys-031115-011319} {\bibfield  {journal}
  {\bibinfo  {journal} {Annu. Rev. Condens. Matter Phys.}\ }\textbf {\bibinfo
  {volume} {7}},\ \bibinfo {pages} {195} (\bibinfo {year} {2016})}\BibitemShut
  {NoStop}%
\bibitem [{\citenamefont {Winter}\ \emph {et~al.}(2017)\citenamefont {Winter},
  \citenamefont {Tsirlin}, \citenamefont {Daghofer}, \citenamefont {{van den
  Brink}}, \citenamefont {Singh}, \citenamefont {Gegenwart},\ and\
  \citenamefont {Valent{\'i}}}]{winterModelsMaterialsGeneralized2017a}%
  \BibitemOpen
  \bibfield  {author} {\bibinfo {author} {\bibfnamefont {S.~M.}\ \bibnamefont
  {Winter}}, \bibinfo {author} {\bibfnamefont {A.~A.}\ \bibnamefont {Tsirlin}},
  \bibinfo {author} {\bibfnamefont {M.}~\bibnamefont {Daghofer}}, \bibinfo
  {author} {\bibfnamefont {J.}~\bibnamefont {{van den Brink}}}, \bibinfo
  {author} {\bibfnamefont {Y.}~\bibnamefont {Singh}}, \bibinfo {author}
  {\bibfnamefont {P.}~\bibnamefont {Gegenwart}}, \ and\ \bibinfo {author}
  {\bibfnamefont {R.}~\bibnamefont {Valent{\'i}}},\ }\href {\doibase
  10.1088/1361-648X/aa8cf5} {\bibfield  {journal} {\bibinfo  {journal} {J.
  Phys.: Condens. Matter}\ }\textbf {\bibinfo {volume} {29}},\ \bibinfo {pages}
  {493002} (\bibinfo {year} {2017})}\BibitemShut {NoStop}%
\bibitem [{\citenamefont {Hermanns}\ \emph {et~al.}(2018)\citenamefont
  {Hermanns}, \citenamefont {Kimchi},\ and\ \citenamefont
  {Knolle}}]{hermannsPhysicsKitaevModel2018}%
  \BibitemOpen
  \bibfield  {author} {\bibinfo {author} {\bibfnamefont {M.}~\bibnamefont
  {Hermanns}}, \bibinfo {author} {\bibfnamefont {I.}~\bibnamefont {Kimchi}}, \
  and\ \bibinfo {author} {\bibfnamefont {J.}~\bibnamefont {Knolle}},\ }\href
  {\doibase 10.1146/annurev-conmatphys-033117-053934} {\bibfield  {journal}
  {\bibinfo  {journal} {Annu. Rev. Condens. Matter Phys.}\ }\textbf {\bibinfo
  {volume} {9}},\ \bibinfo {pages} {17} (\bibinfo {year} {2018})}\BibitemShut
  {NoStop}%
\bibitem [{\citenamefont {Takagi}\ \emph {et~al.}(2019)\citenamefont {Takagi},
  \citenamefont {Takayama}, \citenamefont {Jackeli}, \citenamefont
  {Khaliullin},\ and\ \citenamefont
  {Nagler}}]{takagiConceptRealizationKitaev2019}%
  \BibitemOpen
  \bibfield  {author} {\bibinfo {author} {\bibfnamefont {H.}~\bibnamefont
  {Takagi}}, \bibinfo {author} {\bibfnamefont {T.}~\bibnamefont {Takayama}},
  \bibinfo {author} {\bibfnamefont {G.}~\bibnamefont {Jackeli}}, \bibinfo
  {author} {\bibfnamefont {G.}~\bibnamefont {Khaliullin}}, \ and\ \bibinfo
  {author} {\bibfnamefont {S.~E.}\ \bibnamefont {Nagler}},\ }\href {\doibase
  10.1038/s42254-019-0038-2} {\bibfield  {journal} {\bibinfo  {journal} {Nat
  Rev Phys}\ }\textbf {\bibinfo {volume} {1}},\ \bibinfo {pages} {264}
  (\bibinfo {year} {2019})}\BibitemShut {NoStop}%
\bibitem [{\citenamefont {Motome}\ and\ \citenamefont
  {Nasu}(2020)}]{motomeHuntingMajoranaFermions2020a}%
  \BibitemOpen
  \bibfield  {author} {\bibinfo {author} {\bibfnamefont {Y.}~\bibnamefont
  {Motome}}\ and\ \bibinfo {author} {\bibfnamefont {J.}~\bibnamefont {Nasu}},\
  }\href {\doibase 10.7566/JPSJ.89.012002} {\bibfield  {journal} {\bibinfo
  {journal} {J. Phys. Soc. Jpn.}\ }\textbf {\bibinfo {volume} {89}},\ \bibinfo
  {pages} {012002} (\bibinfo {year} {2020})}\BibitemShut {NoStop}%
\bibitem [{\citenamefont {Trebst}\ and\ \citenamefont
  {Hickey}(2022)}]{trebstKitaevMaterials2022}%
  \BibitemOpen
  \bibfield  {author} {\bibinfo {author} {\bibfnamefont {S.}~\bibnamefont
  {Trebst}}\ and\ \bibinfo {author} {\bibfnamefont {C.}~\bibnamefont
  {Hickey}},\ }\href {\doibase 10.1016/j.physrep.2021.11.003} {\bibfield
  {journal} {\bibinfo  {journal} {Physics Reports}\ }\textbf {\bibinfo {volume}
  {950}},\ \bibinfo {pages} {1} (\bibinfo {year} {2022})}\BibitemShut {NoStop}%
\bibitem [{\citenamefont {Kim}\ \emph {et~al.}(2022)\citenamefont {Kim},
  \citenamefont {Yuan},\ and\ \citenamefont {Kim}}]{kimRuClOtherKitaev2022}%
  \BibitemOpen
  \bibfield  {author} {\bibinfo {author} {\bibfnamefont {S.}~\bibnamefont
  {Kim}}, \bibinfo {author} {\bibfnamefont {B.}~\bibnamefont {Yuan}}, \ and\
  \bibinfo {author} {\bibfnamefont {Y.-J.}\ \bibnamefont {Kim}},\ }\href
  {\doibase 10.1063/5.0101512} {\bibfield  {journal} {\bibinfo  {journal} {APL
  Materials}\ }\textbf {\bibinfo {volume} {10}},\ \bibinfo {pages} {080903}
  (\bibinfo {year} {2022})}\BibitemShut {NoStop}%
\bibitem [{\citenamefont {Chaloupka}\ \emph {et~al.}(2010)\citenamefont
  {Chaloupka}, \citenamefont {Jackeli},\ and\ \citenamefont
  {Khaliullin}}]{chaloupkaKitaevHeisenbergModelHoneycomb2010}%
  \BibitemOpen
  \bibfield  {author} {\bibinfo {author} {\bibfnamefont {J.}~\bibnamefont
  {Chaloupka}}, \bibinfo {author} {\bibfnamefont {G.}~\bibnamefont {Jackeli}},
  \ and\ \bibinfo {author} {\bibfnamefont {G.}~\bibnamefont {Khaliullin}},\
  }\href {\doibase 10.1103/PhysRevLett.105.027204} {\bibfield  {journal}
  {\bibinfo  {journal} {Phys. Rev. Lett.}\ }\textbf {\bibinfo {volume} {105}},\
  \bibinfo {pages} {027204} (\bibinfo {year} {2010})}\BibitemShut {NoStop}%
\bibitem [{\citenamefont {Singh}\ \emph {et~al.}(2012)\citenamefont {Singh},
  \citenamefont {Manni}, \citenamefont {Reuther}, \citenamefont {Berlijn},
  \citenamefont {Thomale}, \citenamefont {Ku}, \citenamefont {Trebst},\ and\
  \citenamefont {Gegenwart}}]{singhRelevanceHeisenbergKitaevModel2012a}%
  \BibitemOpen
  \bibfield  {author} {\bibinfo {author} {\bibfnamefont {Y.}~\bibnamefont
  {Singh}}, \bibinfo {author} {\bibfnamefont {S.}~\bibnamefont {Manni}},
  \bibinfo {author} {\bibfnamefont {J.}~\bibnamefont {Reuther}}, \bibinfo
  {author} {\bibfnamefont {T.}~\bibnamefont {Berlijn}}, \bibinfo {author}
  {\bibfnamefont {R.}~\bibnamefont {Thomale}}, \bibinfo {author} {\bibfnamefont
  {W.}~\bibnamefont {Ku}}, \bibinfo {author} {\bibfnamefont {S.}~\bibnamefont
  {Trebst}}, \ and\ \bibinfo {author} {\bibfnamefont {P.}~\bibnamefont
  {Gegenwart}},\ }\href {\doibase 10.1103/PhysRevLett.108.127203} {\bibfield
  {journal} {\bibinfo  {journal} {Phys. Rev. Lett.}\ }\textbf {\bibinfo
  {volume} {108}},\ \bibinfo {pages} {127203} (\bibinfo {year}
  {2012})}\BibitemShut {NoStop}%
\bibitem [{\citenamefont {Shitade}\ \emph {et~al.}(2009)\citenamefont
  {Shitade}, \citenamefont {Katsura}, \citenamefont {Kune{\v s}}, \citenamefont
  {Qi}, \citenamefont {Zhang},\ and\ \citenamefont
  {Nagaosa}}]{shitadeQuantumSpinHall2009a}%
  \BibitemOpen
  \bibfield  {author} {\bibinfo {author} {\bibfnamefont {A.}~\bibnamefont
  {Shitade}}, \bibinfo {author} {\bibfnamefont {H.}~\bibnamefont {Katsura}},
  \bibinfo {author} {\bibfnamefont {J.}~\bibnamefont {Kune{\v s}}}, \bibinfo
  {author} {\bibfnamefont {X.-L.}\ \bibnamefont {Qi}}, \bibinfo {author}
  {\bibfnamefont {S.-C.}\ \bibnamefont {Zhang}}, \ and\ \bibinfo {author}
  {\bibfnamefont {N.}~\bibnamefont {Nagaosa}},\ }\href {\doibase
  10.1103/PhysRevLett.102.256403} {\bibfield  {journal} {\bibinfo  {journal}
  {Phys. Rev. Lett.}\ }\textbf {\bibinfo {volume} {102}},\ \bibinfo {pages}
  {256403} (\bibinfo {year} {2009})}\BibitemShut {NoStop}%
\bibitem [{\citenamefont {Singh}\ and\ \citenamefont
  {Gegenwart}(2010)}]{singhAntiferromagneticMottInsulating2010b}%
  \BibitemOpen
  \bibfield  {author} {\bibinfo {author} {\bibfnamefont {Y.}~\bibnamefont
  {Singh}}\ and\ \bibinfo {author} {\bibfnamefont {P.}~\bibnamefont
  {Gegenwart}},\ }\href {\doibase 10.1103/PhysRevB.82.064412} {\bibfield
  {journal} {\bibinfo  {journal} {Phys. Rev. B}\ }\textbf {\bibinfo {volume}
  {82}},\ \bibinfo {pages} {064412} (\bibinfo {year} {2010})}\BibitemShut
  {NoStop}%
\bibitem [{\citenamefont {Chaloupka}\ \emph {et~al.}(2013)\citenamefont
  {Chaloupka}, \citenamefont {Jackeli},\ and\ \citenamefont
  {Khaliullin}}]{chaloupkaZigzagMagneticOrder2013}%
  \BibitemOpen
  \bibfield  {author} {\bibinfo {author} {\bibfnamefont {J.}~\bibnamefont
  {Chaloupka}}, \bibinfo {author} {\bibfnamefont {G.}~\bibnamefont {Jackeli}},
  \ and\ \bibinfo {author} {\bibfnamefont {G.}~\bibnamefont {Khaliullin}},\
  }\href {\doibase 10.1103/PhysRevLett.110.097204} {\bibfield  {journal}
  {\bibinfo  {journal} {Phys. Rev. Lett.}\ }\textbf {\bibinfo {volume} {110}},\
  \bibinfo {pages} {097204} (\bibinfo {year} {2013})}\BibitemShut {NoStop}%
\bibitem [{\citenamefont {Reuther}\ \emph {et~al.}(2011)\citenamefont
  {Reuther}, \citenamefont {Thomale},\ and\ \citenamefont
  {Trebst}}]{reutherFinitetemperaturePhaseDiagram2011}%
  \BibitemOpen
  \bibfield  {author} {\bibinfo {author} {\bibfnamefont {J.}~\bibnamefont
  {Reuther}}, \bibinfo {author} {\bibfnamefont {R.}~\bibnamefont {Thomale}}, \
  and\ \bibinfo {author} {\bibfnamefont {S.}~\bibnamefont {Trebst}},\ }\href
  {\doibase 10.1103/PhysRevB.84.100406} {\bibfield  {journal} {\bibinfo
  {journal} {Phys. Rev. B}\ }\textbf {\bibinfo {volume} {84}},\ \bibinfo
  {pages} {100406} (\bibinfo {year} {2011})}\BibitemShut {NoStop}%
\bibitem [{\citenamefont {Mazin}\ \emph {et~al.}(2012)\citenamefont {Mazin},
  \citenamefont {Jeschke}, \citenamefont {Foyevtsova}, \citenamefont
  {Valent{\'i}},\ and\ \citenamefont {Khomskii}}]{mazinNaIrOMolecular2012}%
  \BibitemOpen
  \bibfield  {author} {\bibinfo {author} {\bibfnamefont {I.~I.}\ \bibnamefont
  {Mazin}}, \bibinfo {author} {\bibfnamefont {H.~O.}\ \bibnamefont {Jeschke}},
  \bibinfo {author} {\bibfnamefont {K.}~\bibnamefont {Foyevtsova}}, \bibinfo
  {author} {\bibfnamefont {R.}~\bibnamefont {Valent{\'i}}}, \ and\ \bibinfo
  {author} {\bibfnamefont {D.~I.}\ \bibnamefont {Khomskii}},\ }\href {\doibase
  10.1103/PhysRevLett.109.197201} {\bibfield  {journal} {\bibinfo  {journal}
  {Phys. Rev. Lett.}\ }\textbf {\bibinfo {volume} {109}},\ \bibinfo {pages}
  {197201} (\bibinfo {year} {2012})}\BibitemShut {NoStop}%
\bibitem [{\citenamefont {Plumb}\ \emph {et~al.}(2014)\citenamefont {Plumb},
  \citenamefont {Clancy}, \citenamefont {Sandilands}, \citenamefont {Shankar},
  \citenamefont {Hu}, \citenamefont {Burch}, \citenamefont {Kee},\ and\
  \citenamefont {Kim}}]{plumbRuClSpinorbitAssisted2014}%
  \BibitemOpen
  \bibfield  {author} {\bibinfo {author} {\bibfnamefont {K.~W.}\ \bibnamefont
  {Plumb}}, \bibinfo {author} {\bibfnamefont {J.~P.}\ \bibnamefont {Clancy}},
  \bibinfo {author} {\bibfnamefont {L.~J.}\ \bibnamefont {Sandilands}},
  \bibinfo {author} {\bibfnamefont {V.~V.}\ \bibnamefont {Shankar}}, \bibinfo
  {author} {\bibfnamefont {Y.~F.}\ \bibnamefont {Hu}}, \bibinfo {author}
  {\bibfnamefont {K.~S.}\ \bibnamefont {Burch}}, \bibinfo {author}
  {\bibfnamefont {H.-Y.}\ \bibnamefont {Kee}}, \ and\ \bibinfo {author}
  {\bibfnamefont {Y.-J.}\ \bibnamefont {Kim}},\ }\href {\doibase
  10.1103/PhysRevB.90.041112} {\bibfield  {journal} {\bibinfo  {journal} {Phys.
  Rev. B}\ }\textbf {\bibinfo {volume} {90}},\ \bibinfo {pages} {041112}
  (\bibinfo {year} {2014})}\BibitemShut {NoStop}%
\bibitem [{\citenamefont {Sears}\ \emph {et~al.}(2015)\citenamefont {Sears},
  \citenamefont {Songvilay}, \citenamefont {Plumb}, \citenamefont {Clancy},
  \citenamefont {Qiu}, \citenamefont {Zhao}, \citenamefont {Parshall},\ and\
  \citenamefont {Kim}}]{searsMagneticOrderRuCl2015}%
  \BibitemOpen
  \bibfield  {author} {\bibinfo {author} {\bibfnamefont {J.~A.}\ \bibnamefont
  {Sears}}, \bibinfo {author} {\bibfnamefont {M.}~\bibnamefont {Songvilay}},
  \bibinfo {author} {\bibfnamefont {K.~W.}\ \bibnamefont {Plumb}}, \bibinfo
  {author} {\bibfnamefont {J.~P.}\ \bibnamefont {Clancy}}, \bibinfo {author}
  {\bibfnamefont {Y.}~\bibnamefont {Qiu}}, \bibinfo {author} {\bibfnamefont
  {Y.}~\bibnamefont {Zhao}}, \bibinfo {author} {\bibfnamefont {D.}~\bibnamefont
  {Parshall}}, \ and\ \bibinfo {author} {\bibfnamefont {Y.-J.}\ \bibnamefont
  {Kim}},\ }\href {\doibase 10.1103/PhysRevB.91.144420} {\bibfield  {journal}
  {\bibinfo  {journal} {Phys. Rev. B}\ }\textbf {\bibinfo {volume} {91}},\
  \bibinfo {pages} {144420} (\bibinfo {year} {2015})}\BibitemShut {NoStop}%
\bibitem [{\citenamefont {Banerjee}\ \emph {et~al.}(2016)\citenamefont
  {Banerjee}, \citenamefont {Bridges}, \citenamefont {Yan}, \citenamefont
  {Aczel}, \citenamefont {Li}, \citenamefont {Stone}, \citenamefont {Granroth},
  \citenamefont {Lumsden}, \citenamefont {Yiu}, \citenamefont {Knolle},
  \citenamefont {Bhattacharjee}, \citenamefont {Kovrizhin}, \citenamefont
  {Moessner}, \citenamefont {Tennant}, \citenamefont {Mandrus},\ and\
  \citenamefont {Nagler}}]{banerjeeProximateKitaevQuantum2016a}%
  \BibitemOpen
  \bibfield  {author} {\bibinfo {author} {\bibfnamefont {A.}~\bibnamefont
  {Banerjee}}, \bibinfo {author} {\bibfnamefont {C.~A.}\ \bibnamefont
  {Bridges}}, \bibinfo {author} {\bibfnamefont {J.-Q.}\ \bibnamefont {Yan}},
  \bibinfo {author} {\bibfnamefont {A.~A.}\ \bibnamefont {Aczel}}, \bibinfo
  {author} {\bibfnamefont {L.}~\bibnamefont {Li}}, \bibinfo {author}
  {\bibfnamefont {M.~B.}\ \bibnamefont {Stone}}, \bibinfo {author}
  {\bibfnamefont {G.~E.}\ \bibnamefont {Granroth}}, \bibinfo {author}
  {\bibfnamefont {M.~D.}\ \bibnamefont {Lumsden}}, \bibinfo {author}
  {\bibfnamefont {Y.}~\bibnamefont {Yiu}}, \bibinfo {author} {\bibfnamefont
  {J.}~\bibnamefont {Knolle}}, \bibinfo {author} {\bibfnamefont
  {S.}~\bibnamefont {Bhattacharjee}}, \bibinfo {author} {\bibfnamefont {D.~L.}\
  \bibnamefont {Kovrizhin}}, \bibinfo {author} {\bibfnamefont {R.}~\bibnamefont
  {Moessner}}, \bibinfo {author} {\bibfnamefont {D.~A.}\ \bibnamefont
  {Tennant}}, \bibinfo {author} {\bibfnamefont {D.~G.}\ \bibnamefont
  {Mandrus}}, \ and\ \bibinfo {author} {\bibfnamefont {S.~E.}\ \bibnamefont
  {Nagler}},\ }\href {\doibase 10.1038/nmat4604} {\bibfield  {journal}
  {\bibinfo  {journal} {Nature Mater}\ }\textbf {\bibinfo {volume} {15}},\
  \bibinfo {pages} {733} (\bibinfo {year} {2016})}\BibitemShut {NoStop}%
\bibitem [{\citenamefont {Do}\ \emph {et~al.}(2017)\citenamefont {Do},
  \citenamefont {Park}, \citenamefont {Yoshitake}, \citenamefont {Nasu},
  \citenamefont {Motome}, \citenamefont {Kwon}, \citenamefont {Adroja},
  \citenamefont {Voneshen}, \citenamefont {Kim}, \citenamefont {Jang},
  \citenamefont {Park}, \citenamefont {Choi},\ and\ \citenamefont
  {Ji}}]{doMajoranaFermionsKitaev2017}%
  \BibitemOpen
  \bibfield  {author} {\bibinfo {author} {\bibfnamefont {S.-H.}\ \bibnamefont
  {Do}}, \bibinfo {author} {\bibfnamefont {S.-Y.}\ \bibnamefont {Park}},
  \bibinfo {author} {\bibfnamefont {J.}~\bibnamefont {Yoshitake}}, \bibinfo
  {author} {\bibfnamefont {J.}~\bibnamefont {Nasu}}, \bibinfo {author}
  {\bibfnamefont {Y.}~\bibnamefont {Motome}}, \bibinfo {author} {\bibfnamefont
  {Y.~S.}\ \bibnamefont {Kwon}}, \bibinfo {author} {\bibfnamefont {D.~T.}\
  \bibnamefont {Adroja}}, \bibinfo {author} {\bibfnamefont {D.~J.}\
  \bibnamefont {Voneshen}}, \bibinfo {author} {\bibfnamefont {K.}~\bibnamefont
  {Kim}}, \bibinfo {author} {\bibfnamefont {T.-H.}\ \bibnamefont {Jang}},
  \bibinfo {author} {\bibfnamefont {J.-H.}\ \bibnamefont {Park}}, \bibinfo
  {author} {\bibfnamefont {K.-Y.}\ \bibnamefont {Choi}}, \ and\ \bibinfo
  {author} {\bibfnamefont {S.}~\bibnamefont {Ji}},\ }\href {\doibase
  10.1038/nphys4264} {\bibfield  {journal} {\bibinfo  {journal} {Nature Phys}\
  }\textbf {\bibinfo {volume} {13}},\ \bibinfo {pages} {1079} (\bibinfo {year}
  {2017})}\BibitemShut {NoStop}%
\bibitem [{\citenamefont {Sandilands}\ \emph {et~al.}(2015)\citenamefont
  {Sandilands}, \citenamefont {Tian}, \citenamefont {Plumb}, \citenamefont
  {Kim},\ and\ \citenamefont
  {Burch}}]{sandilandsScatteringContinuumPossible2015}%
  \BibitemOpen
  \bibfield  {author} {\bibinfo {author} {\bibfnamefont {L.~J.}\ \bibnamefont
  {Sandilands}}, \bibinfo {author} {\bibfnamefont {Y.}~\bibnamefont {Tian}},
  \bibinfo {author} {\bibfnamefont {K.~W.}\ \bibnamefont {Plumb}}, \bibinfo
  {author} {\bibfnamefont {Y.-J.}\ \bibnamefont {Kim}}, \ and\ \bibinfo
  {author} {\bibfnamefont {K.~S.}\ \bibnamefont {Burch}},\ }\href {\doibase
  10.1103/PhysRevLett.114.147201} {\bibfield  {journal} {\bibinfo  {journal}
  {Phys. Rev. Lett.}\ }\textbf {\bibinfo {volume} {114}},\ \bibinfo {pages}
  {147201} (\bibinfo {year} {2015})}\BibitemShut {NoStop}%
\bibitem [{\citenamefont {O'Malley}\ \emph {et~al.}(2008)\citenamefont
  {O'Malley}, \citenamefont {Verweij},\ and\ \citenamefont
  {Woodward}}]{omalleyStructurePropertiesOrdered2008}%
  \BibitemOpen
  \bibfield  {author} {\bibinfo {author} {\bibfnamefont {M.~J.}\ \bibnamefont
  {O'Malley}}, \bibinfo {author} {\bibfnamefont {H.}~\bibnamefont {Verweij}}, \
  and\ \bibinfo {author} {\bibfnamefont {P.~M.}\ \bibnamefont {Woodward}},\
  }\href {\doibase 10.1016/j.jssc.2008.04.005} {\bibfield  {journal} {\bibinfo
  {journal} {Journal of Solid State Chemistry}\ }\textbf {\bibinfo {volume}
  {181}},\ \bibinfo {pages} {1803} (\bibinfo {year} {2008})}\BibitemShut
  {NoStop}%
\bibitem [{\citenamefont {Stroganov}\ and\ \citenamefont
  {Ovchinnikov}(1957)}]{stroganov1957}%
  \BibitemOpen
  \bibfield  {author} {\bibinfo {author} {\bibfnamefont {E.~V.}\ \bibnamefont
  {Stroganov}}\ and\ \bibinfo {author} {\bibfnamefont {K.~V.}\ \bibnamefont
  {Ovchinnikov}},\ }\href@noop {} {\bibfield  {journal} {\bibinfo  {journal}
  {Ser. Fiz. i Khim.}\ }\textbf {\bibinfo {volume} {12}} (\bibinfo {year}
  {1957})}\BibitemShut {NoStop}%
\bibitem [{\citenamefont {Johnson}\ \emph {et~al.}(2015)\citenamefont
  {Johnson}, \citenamefont {Williams}, \citenamefont {Haghighirad},
  \citenamefont {Singleton}, \citenamefont {Zapf}, \citenamefont {Manuel},
  \citenamefont {Mazin}, \citenamefont {Li}, \citenamefont {Jeschke},
  \citenamefont {Valent{\'i}},\ and\ \citenamefont
  {Coldea}}]{johnsonMonoclinicCrystalStructure2015a}%
  \BibitemOpen
  \bibfield  {author} {\bibinfo {author} {\bibfnamefont {R.~D.}\ \bibnamefont
  {Johnson}}, \bibinfo {author} {\bibfnamefont {S.~C.}\ \bibnamefont
  {Williams}}, \bibinfo {author} {\bibfnamefont {A.~A.}\ \bibnamefont
  {Haghighirad}}, \bibinfo {author} {\bibfnamefont {J.}~\bibnamefont
  {Singleton}}, \bibinfo {author} {\bibfnamefont {V.}~\bibnamefont {Zapf}},
  \bibinfo {author} {\bibfnamefont {P.}~\bibnamefont {Manuel}}, \bibinfo
  {author} {\bibfnamefont {I.~I.}\ \bibnamefont {Mazin}}, \bibinfo {author}
  {\bibfnamefont {Y.}~\bibnamefont {Li}}, \bibinfo {author} {\bibfnamefont
  {H.~O.}\ \bibnamefont {Jeschke}}, \bibinfo {author} {\bibfnamefont
  {R.}~\bibnamefont {Valent{\'i}}}, \ and\ \bibinfo {author} {\bibfnamefont
  {R.}~\bibnamefont {Coldea}},\ }\href {\doibase 10.1103/PhysRevB.92.235119}
  {\bibfield  {journal} {\bibinfo  {journal} {Phys. Rev. B}\ }\textbf {\bibinfo
  {volume} {92}},\ \bibinfo {pages} {235119} (\bibinfo {year}
  {2015})}\BibitemShut {NoStop}%
\bibitem [{\citenamefont {Cao}\ \emph {et~al.}(2016)\citenamefont {Cao},
  \citenamefont {Banerjee}, \citenamefont {Yan}, \citenamefont {Bridges},
  \citenamefont {Lumsden}, \citenamefont {Mandrus}, \citenamefont {Tennant},
  \citenamefont {Chakoumakos},\ and\ \citenamefont
  {Nagler}}]{caoLowtemperatureCrystalMagnetic2016}%
  \BibitemOpen
  \bibfield  {author} {\bibinfo {author} {\bibfnamefont {H.~B.}\ \bibnamefont
  {Cao}}, \bibinfo {author} {\bibfnamefont {A.}~\bibnamefont {Banerjee}},
  \bibinfo {author} {\bibfnamefont {J.-Q.}\ \bibnamefont {Yan}}, \bibinfo
  {author} {\bibfnamefont {C.~A.}\ \bibnamefont {Bridges}}, \bibinfo {author}
  {\bibfnamefont {M.~D.}\ \bibnamefont {Lumsden}}, \bibinfo {author}
  {\bibfnamefont {D.~G.}\ \bibnamefont {Mandrus}}, \bibinfo {author}
  {\bibfnamefont {D.~A.}\ \bibnamefont {Tennant}}, \bibinfo {author}
  {\bibfnamefont {B.~C.}\ \bibnamefont {Chakoumakos}}, \ and\ \bibinfo {author}
  {\bibfnamefont {S.~E.}\ \bibnamefont {Nagler}},\ }\href {\doibase
  10.1103/PhysRevB.93.134423} {\bibfield  {journal} {\bibinfo  {journal} {Phys.
  Rev. B}\ }\textbf {\bibinfo {volume} {93}},\ \bibinfo {pages} {134423}
  (\bibinfo {year} {2016})}\BibitemShut {NoStop}%
\bibitem [{\citenamefont {Rau}\ \emph {et~al.}(2014)\citenamefont {Rau},
  \citenamefont {Lee},\ and\ \citenamefont {Kee}}]{Rau2014}%
  \BibitemOpen
  \bibfield  {author} {\bibinfo {author} {\bibfnamefont {J.~G.}\ \bibnamefont
  {Rau}}, \bibinfo {author} {\bibfnamefont {E.~K.-H.}\ \bibnamefont {Lee}}, \
  and\ \bibinfo {author} {\bibfnamefont {H.-Y.}\ \bibnamefont {Kee}},\ }\href
  {\doibase 10.1103/PhysRevLett.112.077204} {\bibfield  {journal} {\bibinfo
  {journal} {Phys. Rev. Lett.}\ }\textbf {\bibinfo {volume} {112}},\ \bibinfo
  {pages} {077204} (\bibinfo {year} {2014})}\BibitemShut {NoStop}%
\bibitem [{\citenamefont {Katukuri}\ \emph {et~al.}(2014)\citenamefont
  {Katukuri}, \citenamefont {Nishimoto}, \citenamefont {Yushankhai},
  \citenamefont {Stoyanova}, \citenamefont {Kandpal}, \citenamefont {Choi},
  \citenamefont {Coldea}, \citenamefont {Rousochatzakis}, \citenamefont
  {Hozoi},\ and\ \citenamefont {van~den Brink}}]{Katukuri2014}%
  \BibitemOpen
  \bibfield  {author} {\bibinfo {author} {\bibfnamefont {V.~M.}\ \bibnamefont
  {Katukuri}}, \bibinfo {author} {\bibfnamefont {S.}~\bibnamefont {Nishimoto}},
  \bibinfo {author} {\bibfnamefont {V.}~\bibnamefont {Yushankhai}}, \bibinfo
  {author} {\bibfnamefont {A.}~\bibnamefont {Stoyanova}}, \bibinfo {author}
  {\bibfnamefont {H.}~\bibnamefont {Kandpal}}, \bibinfo {author} {\bibfnamefont
  {S.}~\bibnamefont {Choi}}, \bibinfo {author} {\bibfnamefont {R.}~\bibnamefont
  {Coldea}}, \bibinfo {author} {\bibfnamefont {I.}~\bibnamefont
  {Rousochatzakis}}, \bibinfo {author} {\bibfnamefont {L.}~\bibnamefont
  {Hozoi}}, \ and\ \bibinfo {author} {\bibfnamefont {J.}~\bibnamefont {van~den
  Brink}},\ }\href {\doibase 10.1088/1367-2630/16/1/013056} {\bibfield
  {journal} {\bibinfo  {journal} {New Journal of Physics}\ }\textbf {\bibinfo
  {volume} {16}},\ \bibinfo {pages} {013056} (\bibinfo {year}
  {2014})}\BibitemShut {NoStop}%
\bibitem [{\citenamefont {Chaloupka}\ and\ \citenamefont
  {Khaliullin}(2015)}]{Chaloupka2015}%
  \BibitemOpen
  \bibfield  {author} {\bibinfo {author} {\bibfnamefont {J.~c.~v.}\
  \bibnamefont {Chaloupka}}\ and\ \bibinfo {author} {\bibfnamefont
  {G.}~\bibnamefont {Khaliullin}},\ }\href {\doibase
  10.1103/PhysRevB.92.024413} {\bibfield  {journal} {\bibinfo  {journal} {Phys.
  Rev. B}\ }\textbf {\bibinfo {volume} {92}},\ \bibinfo {pages} {024413}
  (\bibinfo {year} {2015})}\BibitemShut {NoStop}%
\bibitem [{\citenamefont {Bhattacharjee}\ \emph {et~al.}(2012)\citenamefont
  {Bhattacharjee}, \citenamefont {Lee},\ and\ \citenamefont
  {Kim}}]{bhattacharjeeSpinOrbitalLocking2012a}%
  \BibitemOpen
  \bibfield  {author} {\bibinfo {author} {\bibfnamefont {S.}~\bibnamefont
  {Bhattacharjee}}, \bibinfo {author} {\bibfnamefont {S.-S.}\ \bibnamefont
  {Lee}}, \ and\ \bibinfo {author} {\bibfnamefont {Y.~B.}\ \bibnamefont
  {Kim}},\ }\href {\doibase 10.1088/1367-2630/14/7/073015} {\bibfield
  {journal} {\bibinfo  {journal} {New J. Phys.}\ }\textbf {\bibinfo {volume}
  {14}},\ \bibinfo {pages} {073015} (\bibinfo {year} {2012})}\BibitemShut
  {NoStop}%
\bibitem [{\citenamefont {Winter}\ \emph {et~al.}(2016)\citenamefont {Winter},
  \citenamefont {Li}, \citenamefont {Jeschke},\ and\ \citenamefont
  {Valent\'{\i}}}]{Winter2016}%
  \BibitemOpen
  \bibfield  {author} {\bibinfo {author} {\bibfnamefont {S.~M.}\ \bibnamefont
  {Winter}}, \bibinfo {author} {\bibfnamefont {Y.}~\bibnamefont {Li}}, \bibinfo
  {author} {\bibfnamefont {H.~O.}\ \bibnamefont {Jeschke}}, \ and\ \bibinfo
  {author} {\bibfnamefont {R.}~\bibnamefont {Valent\'{\i}}},\ }\href {\doibase
  10.1103/PhysRevB.93.214431} {\bibfield  {journal} {\bibinfo  {journal} {Phys.
  Rev. B}\ }\textbf {\bibinfo {volume} {93}},\ \bibinfo {pages} {214431}
  (\bibinfo {year} {2016})}\BibitemShut {NoStop}%
\bibitem [{\citenamefont {Baez}(2019)}]{Baez2019}%
  \BibitemOpen
  \bibfield  {author} {\bibinfo {author} {\bibfnamefont {M.~L.}\ \bibnamefont
  {Baez}},\ }\href {\doibase 10.1103/PhysRevB.99.184436} {\bibfield  {journal}
  {\bibinfo  {journal} {Phys. Rev. B}\ }\textbf {\bibinfo {volume} {99}},\
  \bibinfo {pages} {184436} (\bibinfo {year} {2019})}\BibitemShut {NoStop}%
\bibitem [{\citenamefont {Kimchi}\ \emph {et~al.}(2015)\citenamefont {Kimchi},
  \citenamefont {Coldea},\ and\ \citenamefont {Vishwanath}}]{Kimchi2015}%
  \BibitemOpen
  \bibfield  {author} {\bibinfo {author} {\bibfnamefont {I.}~\bibnamefont
  {Kimchi}}, \bibinfo {author} {\bibfnamefont {R.}~\bibnamefont {Coldea}}, \
  and\ \bibinfo {author} {\bibfnamefont {A.}~\bibnamefont {Vishwanath}},\
  }\href {\doibase 10.1103/PhysRevB.91.245134} {\bibfield  {journal} {\bibinfo
  {journal} {Phys. Rev. B}\ }\textbf {\bibinfo {volume} {91}},\ \bibinfo
  {pages} {245134} (\bibinfo {year} {2015})}\BibitemShut {NoStop}%
\bibitem [{\citenamefont {Lee}\ \emph {et~al.}(2016)\citenamefont {Lee},
  \citenamefont {Rau},\ and\ \citenamefont {Kim}}]{Lee2016}%
  \BibitemOpen
  \bibfield  {author} {\bibinfo {author} {\bibfnamefont {E.~K.-H.}\
  \bibnamefont {Lee}}, \bibinfo {author} {\bibfnamefont {J.~G.}\ \bibnamefont
  {Rau}}, \ and\ \bibinfo {author} {\bibfnamefont {Y.~B.}\ \bibnamefont
  {Kim}},\ }\href {\doibase 10.1103/PhysRevB.93.184420} {\bibfield  {journal}
  {\bibinfo  {journal} {Phys. Rev. B}\ }\textbf {\bibinfo {volume} {93}},\
  \bibinfo {pages} {184420} (\bibinfo {year} {2016})}\BibitemShut {NoStop}%
\bibitem [{\citenamefont {Gordon}\ \emph {et~al.}(2019)\citenamefont {Gordon},
  \citenamefont {Catuneanu}, \citenamefont {Sørensen},\ and\ \citenamefont
  {Kee}}]{Gordon2019}%
  \BibitemOpen
  \bibfield  {author} {\bibinfo {author} {\bibfnamefont {J.~S.}\ \bibnamefont
  {Gordon}}, \bibinfo {author} {\bibfnamefont {A.}~\bibnamefont {Catuneanu}},
  \bibinfo {author} {\bibfnamefont {E.~S.}\ \bibnamefont {Sørensen}}, \ and\
  \bibinfo {author} {\bibfnamefont {H.-Y.}\ \bibnamefont {Kee}},\ }\href
  {\doibase 10.1038/s41467-019-10405-8} {\bibfield  {journal} {\bibinfo
  {journal} {Nat Commun}\ }\textbf {\bibinfo {volume} {10}},\ \bibinfo {pages}
  {2470} (\bibinfo {year} {2019})}\BibitemShut {NoStop}%
\bibitem [{\citenamefont {Sears}\ \emph {et~al.}(2020)\citenamefont {Sears},
  \citenamefont {Chern}, \citenamefont {Kim}, \citenamefont {Bereciartua},
  \citenamefont {Francoual}, \citenamefont {Kim},\ and\ \citenamefont
  {Kim}}]{searsFerromagneticKitaevInteraction2020a}%
  \BibitemOpen
  \bibfield  {author} {\bibinfo {author} {\bibfnamefont {J.~A.}\ \bibnamefont
  {Sears}}, \bibinfo {author} {\bibfnamefont {L.~E.}\ \bibnamefont {Chern}},
  \bibinfo {author} {\bibfnamefont {S.}~\bibnamefont {Kim}}, \bibinfo {author}
  {\bibfnamefont {P.~J.}\ \bibnamefont {Bereciartua}}, \bibinfo {author}
  {\bibfnamefont {S.}~\bibnamefont {Francoual}}, \bibinfo {author}
  {\bibfnamefont {Y.~B.}\ \bibnamefont {Kim}}, \ and\ \bibinfo {author}
  {\bibfnamefont {Y.-J.}\ \bibnamefont {Kim}},\ }\href {\doibase
  10.1038/s41567-020-0874-0} {\bibfield  {journal} {\bibinfo  {journal} {Nat.
  Phys.}\ }\textbf {\bibinfo {volume} {16}},\ \bibinfo {pages} {837} (\bibinfo
  {year} {2020})}\BibitemShut {NoStop}%
\bibitem [{\citenamefont {Foyevtsova}\ \emph {et~al.}(2013)\citenamefont
  {Foyevtsova}, \citenamefont {Jeschke}, \citenamefont {Mazin}, \citenamefont
  {Khomskii},\ and\ \citenamefont {Valent\'{\i}}}]{Foyevtsova2013}%
  \BibitemOpen
  \bibfield  {author} {\bibinfo {author} {\bibfnamefont {K.}~\bibnamefont
  {Foyevtsova}}, \bibinfo {author} {\bibfnamefont {H.~O.}\ \bibnamefont
  {Jeschke}}, \bibinfo {author} {\bibfnamefont {I.~I.}\ \bibnamefont {Mazin}},
  \bibinfo {author} {\bibfnamefont {D.~I.}\ \bibnamefont {Khomskii}}, \ and\
  \bibinfo {author} {\bibfnamefont {R.}~\bibnamefont {Valent\'{\i}}},\ }\href
  {\doibase 10.1103/PhysRevB.88.035107} {\bibfield  {journal} {\bibinfo
  {journal} {Phys. Rev. B}\ }\textbf {\bibinfo {volume} {88}},\ \bibinfo
  {pages} {035107} (\bibinfo {year} {2013})}\BibitemShut {NoStop}%
\bibitem [{\citenamefont {Gretarsson}\ \emph
  {et~al.}(2013{\natexlab{a}})\citenamefont {Gretarsson}, \citenamefont
  {Clancy}, \citenamefont {Liu}, \citenamefont {Hill}, \citenamefont {Bozin},
  \citenamefont {Singh}, \citenamefont {Manni}, \citenamefont {Gegenwart},
  \citenamefont {Kim}, \citenamefont {Said}, \citenamefont {Casa},
  \citenamefont {Gog}, \citenamefont {Upton}, \citenamefont {Kim},
  \citenamefont {Yu}, \citenamefont {Katukuri}, \citenamefont {Hozoi},
  \citenamefont {{van den Brink}},\ and\ \citenamefont
  {Kim}}]{gretarssonCrystalFieldSplittingCorrelation2013}%
  \BibitemOpen
  \bibfield  {author} {\bibinfo {author} {\bibfnamefont {H.}~\bibnamefont
  {Gretarsson}}, \bibinfo {author} {\bibfnamefont {J.~P.}\ \bibnamefont
  {Clancy}}, \bibinfo {author} {\bibfnamefont {X.}~\bibnamefont {Liu}},
  \bibinfo {author} {\bibfnamefont {J.~P.}\ \bibnamefont {Hill}}, \bibinfo
  {author} {\bibfnamefont {E.}~\bibnamefont {Bozin}}, \bibinfo {author}
  {\bibfnamefont {Y.}~\bibnamefont {Singh}}, \bibinfo {author} {\bibfnamefont
  {S.}~\bibnamefont {Manni}}, \bibinfo {author} {\bibfnamefont
  {P.}~\bibnamefont {Gegenwart}}, \bibinfo {author} {\bibfnamefont
  {J.}~\bibnamefont {Kim}}, \bibinfo {author} {\bibfnamefont {A.~H.}\
  \bibnamefont {Said}}, \bibinfo {author} {\bibfnamefont {D.}~\bibnamefont
  {Casa}}, \bibinfo {author} {\bibfnamefont {T.}~\bibnamefont {Gog}}, \bibinfo
  {author} {\bibfnamefont {M.~H.}\ \bibnamefont {Upton}}, \bibinfo {author}
  {\bibfnamefont {H.-S.}\ \bibnamefont {Kim}}, \bibinfo {author} {\bibfnamefont
  {J.}~\bibnamefont {Yu}}, \bibinfo {author} {\bibfnamefont {V.~M.}\
  \bibnamefont {Katukuri}}, \bibinfo {author} {\bibfnamefont {L.}~\bibnamefont
  {Hozoi}}, \bibinfo {author} {\bibfnamefont {J.}~\bibnamefont {{van den
  Brink}}}, \ and\ \bibinfo {author} {\bibfnamefont {Y.-J.}\ \bibnamefont
  {Kim}},\ }\href {\doibase 10.1103/PhysRevLett.110.076402} {\bibfield
  {journal} {\bibinfo  {journal} {Phys. Rev. Lett.}\ }\textbf {\bibinfo
  {volume} {110}},\ \bibinfo {pages} {076402} (\bibinfo {year}
  {2013}{\natexlab{a}})}\BibitemShut {NoStop}%
\bibitem [{\citenamefont {Hermann}\ \emph {et~al.}(2019)\citenamefont
  {Hermann}, \citenamefont {{Ebad-Allah}}, \citenamefont {Freund},
  \citenamefont {Jesche}, \citenamefont {Tsirlin}, \citenamefont {Gegenwart},\
  and\ \citenamefont {Kuntscher}}]{Hermann2019}%
  \BibitemOpen
  \bibfield  {author} {\bibinfo {author} {\bibfnamefont {V.}~\bibnamefont
  {Hermann}}, \bibinfo {author} {\bibfnamefont {J.}~\bibnamefont
  {{Ebad-Allah}}}, \bibinfo {author} {\bibfnamefont {F.}~\bibnamefont
  {Freund}}, \bibinfo {author} {\bibfnamefont {A.}~\bibnamefont {Jesche}},
  \bibinfo {author} {\bibfnamefont {A.~A.}\ \bibnamefont {Tsirlin}}, \bibinfo
  {author} {\bibfnamefont {P.}~\bibnamefont {Gegenwart}}, \ and\ \bibinfo
  {author} {\bibfnamefont {C.~A.}\ \bibnamefont {Kuntscher}},\ }\href {\doibase
  10.1103/PhysRevB.99.235116} {\bibfield  {journal} {\bibinfo  {journal} {Phys.
  Rev. B}\ }\textbf {\bibinfo {volume} {99}},\ \bibinfo {pages} {235116}
  (\bibinfo {year} {2019})}\BibitemShut {NoStop}%
\bibitem [{\citenamefont {Comin}\ \emph {et~al.}(2012)\citenamefont {Comin},
  \citenamefont {Levy}, \citenamefont {Ludbrook}, \citenamefont {Zhu},
  \citenamefont {Veenstra}, \citenamefont {Rosen}, \citenamefont {Singh},
  \citenamefont {Gegenwart}, \citenamefont {Stricker}, \citenamefont {Hancock},
  \citenamefont {{van der Marel}}, \citenamefont {Elfimov},\ and\ \citenamefont
  {Damascelli}}]{Comin2012}%
  \BibitemOpen
  \bibfield  {author} {\bibinfo {author} {\bibfnamefont {R.}~\bibnamefont
  {Comin}}, \bibinfo {author} {\bibfnamefont {G.}~\bibnamefont {Levy}},
  \bibinfo {author} {\bibfnamefont {B.}~\bibnamefont {Ludbrook}}, \bibinfo
  {author} {\bibfnamefont {Z.-H.}\ \bibnamefont {Zhu}}, \bibinfo {author}
  {\bibfnamefont {C.~N.}\ \bibnamefont {Veenstra}}, \bibinfo {author}
  {\bibfnamefont {J.~A.}\ \bibnamefont {Rosen}}, \bibinfo {author}
  {\bibfnamefont {Y.}~\bibnamefont {Singh}}, \bibinfo {author} {\bibfnamefont
  {P.}~\bibnamefont {Gegenwart}}, \bibinfo {author} {\bibfnamefont
  {D.}~\bibnamefont {Stricker}}, \bibinfo {author} {\bibfnamefont {J.~N.}\
  \bibnamefont {Hancock}}, \bibinfo {author} {\bibfnamefont {D.}~\bibnamefont
  {{van der Marel}}}, \bibinfo {author} {\bibfnamefont {I.~S.}\ \bibnamefont
  {Elfimov}}, \ and\ \bibinfo {author} {\bibfnamefont {A.}~\bibnamefont
  {Damascelli}},\ }\href {\doibase 10.1103/PhysRevLett.109.266406} {\bibfield
  {journal} {\bibinfo  {journal} {Phys. Rev. Lett.}\ }\textbf {\bibinfo
  {volume} {109}},\ \bibinfo {pages} {266406} (\bibinfo {year}
  {2012})}\BibitemShut {NoStop}%
\bibitem [{\citenamefont {Kim}\ \emph {et~al.}(2016)\citenamefont {Kim},
  \citenamefont {Shirakawa},\ and\ \citenamefont {Yunoki}}]{BHKim2016}%
  \BibitemOpen
  \bibfield  {author} {\bibinfo {author} {\bibfnamefont {B.~H.}\ \bibnamefont
  {Kim}}, \bibinfo {author} {\bibfnamefont {T.}~\bibnamefont {Shirakawa}}, \
  and\ \bibinfo {author} {\bibfnamefont {S.}~\bibnamefont {Yunoki}},\ }\href
  {\doibase 10.1103/PhysRevLett.117.187201} {\bibfield  {journal} {\bibinfo
  {journal} {Phys. Rev. Lett.}\ }\textbf {\bibinfo {volume} {117}},\ \bibinfo
  {pages} {187201} (\bibinfo {year} {2016})}\BibitemShut {NoStop}%
\bibitem [{\citenamefont {Kim}\ \emph {et~al.}(2014)\citenamefont {Kim},
  \citenamefont {Khaliullin},\ and\ \citenamefont
  {Min}}]{kimElectronicExcitationsEdgeshared2014b}%
  \BibitemOpen
  \bibfield  {author} {\bibinfo {author} {\bibfnamefont {B.~H.}\ \bibnamefont
  {Kim}}, \bibinfo {author} {\bibfnamefont {G.}~\bibnamefont {Khaliullin}}, \
  and\ \bibinfo {author} {\bibfnamefont {B.~I.}\ \bibnamefont {Min}},\ }\href
  {\doibase 10.1103/PhysRevB.89.081109} {\bibfield  {journal} {\bibinfo
  {journal} {Phys. Rev. B}\ }\textbf {\bibinfo {volume} {89}},\ \bibinfo
  {pages} {081109} (\bibinfo {year} {2014})}\BibitemShut {NoStop}%
\bibitem [{\citenamefont {Igarashi}\ and\ \citenamefont
  {Nagao}(2016{\natexlab{a}})}]{igarashiCollectiveExcitationsNa2016a}%
  \BibitemOpen
  \bibfield  {author} {\bibinfo {author} {\bibfnamefont {J.-I.}\ \bibnamefont
  {Igarashi}}\ and\ \bibinfo {author} {\bibfnamefont {T.}~\bibnamefont
  {Nagao}},\ }\href {\doibase 10.1088/0953-8984/28/2/026006} {\bibfield
  {journal} {\bibinfo  {journal} {J. Phys.: Condens. Matter}\ }\textbf
  {\bibinfo {volume} {28}},\ \bibinfo {pages} {026006} (\bibinfo {year}
  {2016}{\natexlab{a}})}\BibitemShut {NoStop}%
\bibitem [{\citenamefont {Igarashi}\ and\ \citenamefont
  {Nagao}(2016{\natexlab{b}})}]{igarashiResonantInelasticXray2016a}%
  \BibitemOpen
  \bibfield  {author} {\bibinfo {author} {\bibfnamefont {J.-i.}\ \bibnamefont
  {Igarashi}}\ and\ \bibinfo {author} {\bibfnamefont {T.}~\bibnamefont
  {Nagao}},\ }\href {\doibase 10.1016/j.elspec.2016.08.003} {\bibfield
  {journal} {\bibinfo  {journal} {Journal of Electron Spectroscopy and Related
  Phenomena}\ }\textbf {\bibinfo {volume} {212}},\ \bibinfo {pages} {44}
  (\bibinfo {year} {2016}{\natexlab{b}})}\BibitemShut {NoStop}%
\bibitem [{\citenamefont {Lin}\ \emph {et~al.}(2012)\citenamefont {Lin},
  \citenamefont {Li}, \citenamefont {Gray},\ and\ \citenamefont
  {Mitchell}}]{linVaporGrowthChemical2012}%
  \BibitemOpen
  \bibfield  {author} {\bibinfo {author} {\bibfnamefont {Q.}~\bibnamefont
  {Lin}}, \bibinfo {author} {\bibfnamefont {Q.}~\bibnamefont {Li}}, \bibinfo
  {author} {\bibfnamefont {K.~E.}\ \bibnamefont {Gray}}, \ and\ \bibinfo
  {author} {\bibfnamefont {J.~F.}\ \bibnamefont {Mitchell}},\ }\href {\doibase
  10.1021/cg201238n} {\bibfield  {journal} {\bibinfo  {journal} {Crystal Growth
  \& Design}\ }\textbf {\bibinfo {volume} {12}},\ \bibinfo {pages} {1232}
  (\bibinfo {year} {2012})}\BibitemShut {NoStop}%
\bibitem [{\citenamefont {Freund}\ \emph {et~al.}(2016)\citenamefont {Freund},
  \citenamefont {Williams}, \citenamefont {Johnson}, \citenamefont {Coldea},
  \citenamefont {Gegenwart},\ and\ \citenamefont
  {Jesche}}]{freundSingleCrystalGrowth2016}%
  \BibitemOpen
  \bibfield  {author} {\bibinfo {author} {\bibfnamefont {F.}~\bibnamefont
  {Freund}}, \bibinfo {author} {\bibfnamefont {S.~C.}\ \bibnamefont
  {Williams}}, \bibinfo {author} {\bibfnamefont {R.~D.}\ \bibnamefont
  {Johnson}}, \bibinfo {author} {\bibfnamefont {R.}~\bibnamefont {Coldea}},
  \bibinfo {author} {\bibfnamefont {P.}~\bibnamefont {Gegenwart}}, \ and\
  \bibinfo {author} {\bibfnamefont {A.}~\bibnamefont {Jesche}},\ }\href
  {\doibase 10.1038/srep35362} {\bibfield  {journal} {\bibinfo  {journal} {Sci
  Rep}\ }\textbf {\bibinfo {volume} {6}},\ \bibinfo {pages} {35362} (\bibinfo
  {year} {2016})}\BibitemShut {NoStop}%
\bibitem [{\citenamefont {Lebert}\ \emph {et~al.}(2022)\citenamefont {Lebert},
  \citenamefont {Kim}, \citenamefont {Prishchenko}, \citenamefont {Tsirlin},
  \citenamefont {Said}, \citenamefont {Alatas},\ and\ \citenamefont
  {Kim}}]{lebertAcousticPhononDispersion2022}%
  \BibitemOpen
  \bibfield  {author} {\bibinfo {author} {\bibfnamefont {B.~W.}\ \bibnamefont
  {Lebert}}, \bibinfo {author} {\bibfnamefont {S.}~\bibnamefont {Kim}},
  \bibinfo {author} {\bibfnamefont {D.~A.}\ \bibnamefont {Prishchenko}},
  \bibinfo {author} {\bibfnamefont {A.~A.}\ \bibnamefont {Tsirlin}}, \bibinfo
  {author} {\bibfnamefont {A.~H.}\ \bibnamefont {Said}}, \bibinfo {author}
  {\bibfnamefont {A.}~\bibnamefont {Alatas}}, \ and\ \bibinfo {author}
  {\bibfnamefont {Y.-J.}\ \bibnamefont {Kim}},\ }\href {\doibase
  10.1103/PhysRevB.106.L041102} {\bibfield  {journal} {\bibinfo  {journal}
  {Phys. Rev. B}\ }\textbf {\bibinfo {volume} {106}},\ \bibinfo {pages}
  {L041102} (\bibinfo {year} {2022})}\BibitemShut {NoStop}%
\bibitem [{\citenamefont {Chun}\ \emph {et~al.}(2021)\citenamefont {Chun},
  \citenamefont {Stavropoulos}, \citenamefont {Kee}, \citenamefont {Sala},
  \citenamefont {Kim}, \citenamefont {Kim}, \citenamefont {Kim}, \citenamefont
  {Mitchell},\ and\ \citenamefont {Kim}}]{chunOpticalMagnonsDominant2021a}%
  \BibitemOpen
  \bibfield  {author} {\bibinfo {author} {\bibfnamefont {S.~H.}\ \bibnamefont
  {Chun}}, \bibinfo {author} {\bibfnamefont {P.~P.}\ \bibnamefont
  {Stavropoulos}}, \bibinfo {author} {\bibfnamefont {H.-Y.}\ \bibnamefont
  {Kee}}, \bibinfo {author} {\bibfnamefont {M.~M.}\ \bibnamefont {Sala}},
  \bibinfo {author} {\bibfnamefont {J.}~\bibnamefont {Kim}}, \bibinfo {author}
  {\bibfnamefont {J.-W.}\ \bibnamefont {Kim}}, \bibinfo {author} {\bibfnamefont
  {B.~J.}\ \bibnamefont {Kim}}, \bibinfo {author} {\bibfnamefont {J.~F.}\
  \bibnamefont {Mitchell}}, \ and\ \bibinfo {author} {\bibfnamefont {Y.-J.}\
  \bibnamefont {Kim}},\ }\href {\doibase 10.1103/PhysRevB.103.L020410}
  {\bibfield  {journal} {\bibinfo  {journal} {Phys. Rev. B}\ }\textbf {\bibinfo
  {volume} {103}},\ \bibinfo {pages} {L020410} (\bibinfo {year}
  {2021})}\BibitemShut {NoStop}%
\bibitem [{Not()}]{NoteX}%
  \BibitemOpen
  \href@noop {} {}\bibinfo {note} {See Supplemental Material at [URL will be
  inserted by publisher] for theoretical RIXS calculation details and
  additional experimental figures.}\BibitemShut {Stop}%
\bibitem [{\citenamefont {Williams}\ \emph {et~al.}(2016)\citenamefont
  {Williams}, \citenamefont {Johnson}, \citenamefont {Freund}, \citenamefont
  {Choi}, \citenamefont {Jesche}, \citenamefont {Kimchi}, \citenamefont
  {Manni}, \citenamefont {Bombardi}, \citenamefont {Manuel}, \citenamefont
  {Gegenwart},\ and\ \citenamefont
  {Coldea}}]{williamsIncommensurateCounterrotatingMagnetic2016a}%
  \BibitemOpen
  \bibfield  {author} {\bibinfo {author} {\bibfnamefont {S.~C.}\ \bibnamefont
  {Williams}}, \bibinfo {author} {\bibfnamefont {R.~D.}\ \bibnamefont
  {Johnson}}, \bibinfo {author} {\bibfnamefont {F.}~\bibnamefont {Freund}},
  \bibinfo {author} {\bibfnamefont {S.}~\bibnamefont {Choi}}, \bibinfo {author}
  {\bibfnamefont {A.}~\bibnamefont {Jesche}}, \bibinfo {author} {\bibfnamefont
  {I.}~\bibnamefont {Kimchi}}, \bibinfo {author} {\bibfnamefont
  {S.}~\bibnamefont {Manni}}, \bibinfo {author} {\bibfnamefont
  {A.}~\bibnamefont {Bombardi}}, \bibinfo {author} {\bibfnamefont
  {P.}~\bibnamefont {Manuel}}, \bibinfo {author} {\bibfnamefont
  {P.}~\bibnamefont {Gegenwart}}, \ and\ \bibinfo {author} {\bibfnamefont
  {R.}~\bibnamefont {Coldea}},\ }\href {\doibase 10.1103/PhysRevB.93.195158}
  {\bibfield  {journal} {\bibinfo  {journal} {Phys. Rev. B}\ }\textbf {\bibinfo
  {volume} {93}},\ \bibinfo {pages} {195158} (\bibinfo {year}
  {2016})}\BibitemShut {NoStop}%
\bibitem [{\citenamefont {Liu}\ \emph {et~al.}(2011)\citenamefont {Liu},
  \citenamefont {Berlijn}, \citenamefont {Yin}, \citenamefont {Ku},
  \citenamefont {Tsvelik}, \citenamefont {Kim}, \citenamefont {Gretarsson},
  \citenamefont {Singh}, \citenamefont {Gegenwart},\ and\ \citenamefont
  {Hill}}]{liuLongrangeMagneticOrdering2011a}%
  \BibitemOpen
  \bibfield  {author} {\bibinfo {author} {\bibfnamefont {X.}~\bibnamefont
  {Liu}}, \bibinfo {author} {\bibfnamefont {T.}~\bibnamefont {Berlijn}},
  \bibinfo {author} {\bibfnamefont {W.-G.}\ \bibnamefont {Yin}}, \bibinfo
  {author} {\bibfnamefont {W.}~\bibnamefont {Ku}}, \bibinfo {author}
  {\bibfnamefont {A.}~\bibnamefont {Tsvelik}}, \bibinfo {author} {\bibfnamefont
  {Y.-J.}\ \bibnamefont {Kim}}, \bibinfo {author} {\bibfnamefont
  {H.}~\bibnamefont {Gretarsson}}, \bibinfo {author} {\bibfnamefont
  {Y.}~\bibnamefont {Singh}}, \bibinfo {author} {\bibfnamefont
  {P.}~\bibnamefont {Gegenwart}}, \ and\ \bibinfo {author} {\bibfnamefont
  {J.~P.}\ \bibnamefont {Hill}},\ }\href {\doibase 10.1103/PhysRevB.83.220403}
  {\bibfield  {journal} {\bibinfo  {journal} {Phys. Rev. B}\ }\textbf {\bibinfo
  {volume} {83}},\ \bibinfo {pages} {220403} (\bibinfo {year}
  {2011})}\BibitemShut {NoStop}%
\bibitem [{\citenamefont {Mehlawat}\ \emph {et~al.}(2017)\citenamefont
  {Mehlawat}, \citenamefont {Thamizhavel},\ and\ \citenamefont
  {Singh}}]{mehlawatHeatCapacityEvidence2017}%
  \BibitemOpen
  \bibfield  {author} {\bibinfo {author} {\bibfnamefont {K.}~\bibnamefont
  {Mehlawat}}, \bibinfo {author} {\bibfnamefont {A.}~\bibnamefont
  {Thamizhavel}}, \ and\ \bibinfo {author} {\bibfnamefont {Y.}~\bibnamefont
  {Singh}},\ }\href {\doibase 10.1103/PhysRevB.95.144406} {\bibfield  {journal}
  {\bibinfo  {journal} {Phys. Rev. B}\ }\textbf {\bibinfo {volume} {95}},\
  \bibinfo {pages} {144406} (\bibinfo {year} {2017})}\BibitemShut {NoStop}%
\bibitem [{\citenamefont {Ye}\ \emph {et~al.}(2012)\citenamefont {Ye},
  \citenamefont {Chi}, \citenamefont {Cao}, \citenamefont {Chakoumakos},
  \citenamefont {{Fernandez-Baca}}, \citenamefont {Custelcean}, \citenamefont
  {Qi}, \citenamefont {Korneta},\ and\ \citenamefont
  {Cao}}]{yeDirectEvidenceZigzag2012a}%
  \BibitemOpen
  \bibfield  {author} {\bibinfo {author} {\bibfnamefont {F.}~\bibnamefont
  {Ye}}, \bibinfo {author} {\bibfnamefont {S.}~\bibnamefont {Chi}}, \bibinfo
  {author} {\bibfnamefont {H.}~\bibnamefont {Cao}}, \bibinfo {author}
  {\bibfnamefont {B.~C.}\ \bibnamefont {Chakoumakos}}, \bibinfo {author}
  {\bibfnamefont {J.~A.}\ \bibnamefont {{Fernandez-Baca}}}, \bibinfo {author}
  {\bibfnamefont {R.}~\bibnamefont {Custelcean}}, \bibinfo {author}
  {\bibfnamefont {T.~F.}\ \bibnamefont {Qi}}, \bibinfo {author} {\bibfnamefont
  {O.~B.}\ \bibnamefont {Korneta}}, \ and\ \bibinfo {author} {\bibfnamefont
  {G.}~\bibnamefont {Cao}},\ }\href {\doibase 10.1103/PhysRevB.85.180403}
  {\bibfield  {journal} {\bibinfo  {journal} {Phys. Rev. B}\ }\textbf {\bibinfo
  {volume} {85}},\ \bibinfo {pages} {180403} (\bibinfo {year}
  {2012})}\BibitemShut {NoStop}%
\bibitem [{\citenamefont {Sears}\ \emph {et~al.}(2017)\citenamefont {Sears},
  \citenamefont {Zhao}, \citenamefont {Xu}, \citenamefont {Lynn},\ and\
  \citenamefont {Kim}}]{searsPhaseDiagramRuCl2017a}%
  \BibitemOpen
  \bibfield  {author} {\bibinfo {author} {\bibfnamefont {J.~A.}\ \bibnamefont
  {Sears}}, \bibinfo {author} {\bibfnamefont {Y.}~\bibnamefont {Zhao}},
  \bibinfo {author} {\bibfnamefont {Z.}~\bibnamefont {Xu}}, \bibinfo {author}
  {\bibfnamefont {J.~W.}\ \bibnamefont {Lynn}}, \ and\ \bibinfo {author}
  {\bibfnamefont {Y.-J.}\ \bibnamefont {Kim}},\ }\href {\doibase
  10.1103/PhysRevB.95.180411} {\bibfield  {journal} {\bibinfo  {journal} {Phys.
  Rev. B}\ }\textbf {\bibinfo {volume} {95}},\ \bibinfo {pages} {180411}
  (\bibinfo {year} {2017})}\BibitemShut {NoStop}%
\bibitem [{\citenamefont {Shvyd'ko}\ \emph {et~al.}(2013)\citenamefont
  {Shvyd'ko}, \citenamefont {Hill}, \citenamefont {Burns}, \citenamefont
  {Coburn}, \citenamefont {Brajuskovic}, \citenamefont {Casa}, \citenamefont
  {Goetze}, \citenamefont {Gog}, \citenamefont {Khachatryan}, \citenamefont
  {Kim}, \citenamefont {Kodituwakku}, \citenamefont {Ramanathan}, \citenamefont
  {Roberts}, \citenamefont {Said}, \citenamefont {Sinn}, \citenamefont {Shu},
  \citenamefont {Stoupin}, \citenamefont {Upton}, \citenamefont {Wieczorek},\
  and\ \citenamefont {Yavas}}]{shvydkoMERIXNextGeneration2013}%
  \BibitemOpen
  \bibfield  {author} {\bibinfo {author} {\bibfnamefont {Y.}~\bibnamefont
  {Shvyd'ko}}, \bibinfo {author} {\bibfnamefont {J.}~\bibnamefont {Hill}},
  \bibinfo {author} {\bibfnamefont {C.}~\bibnamefont {Burns}}, \bibinfo
  {author} {\bibfnamefont {D.}~\bibnamefont {Coburn}}, \bibinfo {author}
  {\bibfnamefont {B.}~\bibnamefont {Brajuskovic}}, \bibinfo {author}
  {\bibfnamefont {D.}~\bibnamefont {Casa}}, \bibinfo {author} {\bibfnamefont
  {K.}~\bibnamefont {Goetze}}, \bibinfo {author} {\bibfnamefont
  {T.}~\bibnamefont {Gog}}, \bibinfo {author} {\bibfnamefont {R.}~\bibnamefont
  {Khachatryan}}, \bibinfo {author} {\bibfnamefont {J.-H.}\ \bibnamefont
  {Kim}}, \bibinfo {author} {\bibfnamefont {C.}~\bibnamefont {Kodituwakku}},
  \bibinfo {author} {\bibfnamefont {M.}~\bibnamefont {Ramanathan}}, \bibinfo
  {author} {\bibfnamefont {T.}~\bibnamefont {Roberts}}, \bibinfo {author}
  {\bibfnamefont {A.}~\bibnamefont {Said}}, \bibinfo {author} {\bibfnamefont
  {H.}~\bibnamefont {Sinn}}, \bibinfo {author} {\bibfnamefont {D.}~\bibnamefont
  {Shu}}, \bibinfo {author} {\bibfnamefont {S.}~\bibnamefont {Stoupin}},
  \bibinfo {author} {\bibfnamefont {M.}~\bibnamefont {Upton}}, \bibinfo
  {author} {\bibfnamefont {M.}~\bibnamefont {Wieczorek}}, \ and\ \bibinfo
  {author} {\bibfnamefont {H.}~\bibnamefont {Yavas}},\ }\href {\doibase
  10.1016/j.elspec.2012.09.003} {\bibfield  {journal} {\bibinfo  {journal}
  {Journal of Electron Spectroscopy and Related Phenomena}\ }\textbf {\bibinfo
  {volume} {188}},\ \bibinfo {pages} {140} (\bibinfo {year}
  {2013})}\BibitemShut {NoStop}%
\bibitem [{\citenamefont {Zhou}\ \emph {et~al.}(2022)\citenamefont {Zhou},
  \citenamefont {Walters}, \citenamefont {{Garcia-Fernandez}}, \citenamefont
  {Rice}, \citenamefont {Hand}, \citenamefont {Nag}, \citenamefont {Li},
  \citenamefont {Agrestini}, \citenamefont {Garland}, \citenamefont {Wang},
  \citenamefont {Alcock}, \citenamefont {Nistea}, \citenamefont {Nutter},
  \citenamefont {Rubies}, \citenamefont {Knap}, \citenamefont {Gaughran},
  \citenamefont {Yuan}, \citenamefont {Chang}, \citenamefont {Emmins},\ and\
  \citenamefont {Howell}}]{zhouI21AdvancedHighresolution2022}%
  \BibitemOpen
  \bibfield  {author} {\bibinfo {author} {\bibfnamefont {K.-J.}\ \bibnamefont
  {Zhou}}, \bibinfo {author} {\bibfnamefont {A.}~\bibnamefont {Walters}},
  \bibinfo {author} {\bibfnamefont {M.}~\bibnamefont {{Garcia-Fernandez}}},
  \bibinfo {author} {\bibfnamefont {T.}~\bibnamefont {Rice}}, \bibinfo {author}
  {\bibfnamefont {M.}~\bibnamefont {Hand}}, \bibinfo {author} {\bibfnamefont
  {A.}~\bibnamefont {Nag}}, \bibinfo {author} {\bibfnamefont {J.}~\bibnamefont
  {Li}}, \bibinfo {author} {\bibfnamefont {S.}~\bibnamefont {Agrestini}},
  \bibinfo {author} {\bibfnamefont {P.}~\bibnamefont {Garland}}, \bibinfo
  {author} {\bibfnamefont {H.}~\bibnamefont {Wang}}, \bibinfo {author}
  {\bibfnamefont {S.}~\bibnamefont {Alcock}}, \bibinfo {author} {\bibfnamefont
  {I.}~\bibnamefont {Nistea}}, \bibinfo {author} {\bibfnamefont
  {B.}~\bibnamefont {Nutter}}, \bibinfo {author} {\bibfnamefont
  {N.}~\bibnamefont {Rubies}}, \bibinfo {author} {\bibfnamefont
  {G.}~\bibnamefont {Knap}}, \bibinfo {author} {\bibfnamefont {M.}~\bibnamefont
  {Gaughran}}, \bibinfo {author} {\bibfnamefont {F.}~\bibnamefont {Yuan}},
  \bibinfo {author} {\bibfnamefont {P.}~\bibnamefont {Chang}}, \bibinfo
  {author} {\bibfnamefont {J.}~\bibnamefont {Emmins}}, \ and\ \bibinfo {author}
  {\bibfnamefont {G.}~\bibnamefont {Howell}},\ }\href {\doibase
  10.1107/S1600577522000601} {\bibfield  {journal} {\bibinfo  {journal} {J
  Synchrotron Rad}\ }\textbf {\bibinfo {volume} {29}},\ \bibinfo {pages} {563}
  (\bibinfo {year} {2022})}\BibitemShut {NoStop}%
\bibitem [{\citenamefont {Lee}\ \emph {et~al.}(2021)\citenamefont {Lee},
  \citenamefont {Choi}, \citenamefont {Do}, \citenamefont {Kim}, \citenamefont
  {Seong},\ and\ \citenamefont {Choi}}]{Lee2021}%
  \BibitemOpen
  \bibfield  {author} {\bibinfo {author} {\bibfnamefont {J.-H.}\ \bibnamefont
  {Lee}}, \bibinfo {author} {\bibfnamefont {Y.}~\bibnamefont {Choi}}, \bibinfo
  {author} {\bibfnamefont {S.-H.}\ \bibnamefont {Do}}, \bibinfo {author}
  {\bibfnamefont {B.~H.}\ \bibnamefont {Kim}}, \bibinfo {author} {\bibfnamefont
  {M.-J.}\ \bibnamefont {Seong}}, \ and\ \bibinfo {author} {\bibfnamefont
  {K.-Y.}\ \bibnamefont {Choi}},\ }\href {\doibase 10.1038/s41535-021-00340-7}
  {\bibfield  {journal} {\bibinfo  {journal} {npj Quantum Mater.}\ }\textbf
  {\bibinfo {volume} {6}},\ \bibinfo {pages} {43} (\bibinfo {year}
  {2021})}\BibitemShut {NoStop}%
\bibitem [{\citenamefont {Ament}\ \emph {et~al.}(2011)\citenamefont {Ament},
  \citenamefont {{van Veenendaal}}, \citenamefont {Devereaux}, \citenamefont
  {Hill},\ and\ \citenamefont {{van den Brink}}}]{Ament2011}%
  \BibitemOpen
  \bibfield  {author} {\bibinfo {author} {\bibfnamefont {L.~J.~P.}\
  \bibnamefont {Ament}}, \bibinfo {author} {\bibfnamefont {M.}~\bibnamefont
  {{van Veenendaal}}}, \bibinfo {author} {\bibfnamefont {T.~P.}\ \bibnamefont
  {Devereaux}}, \bibinfo {author} {\bibfnamefont {J.~P.}\ \bibnamefont {Hill}},
  \ and\ \bibinfo {author} {\bibfnamefont {J.}~\bibnamefont {{van den
  Brink}}},\ }\href {\doibase 10.1103/RevModPhys.83.705} {\bibfield  {journal}
  {\bibinfo  {journal} {Rev. Mod. Phys.}\ }\textbf {\bibinfo {volume} {83}},\
  \bibinfo {pages} {705} (\bibinfo {year} {2011})}\BibitemShut {NoStop}%
\bibitem [{\citenamefont {Vale}\ \emph {et~al.}(2019)\citenamefont {Vale},
  \citenamefont {Dashwood}, \citenamefont {Paris}, \citenamefont {Veiga},
  \citenamefont {{Garcia-Fernandez}}, \citenamefont {Nag}, \citenamefont
  {Walters}, \citenamefont {Zhou}, \citenamefont {Pietsch}, \citenamefont
  {Jesche}, \citenamefont {Gegenwart}, \citenamefont {Coldea}, \citenamefont
  {Schmitt},\ and\ \citenamefont
  {McMorrow}}]{valeHighresolutionResonantInelastic2019b}%
  \BibitemOpen
  \bibfield  {author} {\bibinfo {author} {\bibfnamefont {J.~G.}\ \bibnamefont
  {Vale}}, \bibinfo {author} {\bibfnamefont {C.~D.}\ \bibnamefont {Dashwood}},
  \bibinfo {author} {\bibfnamefont {E.}~\bibnamefont {Paris}}, \bibinfo
  {author} {\bibfnamefont {L.~S.~I.}\ \bibnamefont {Veiga}}, \bibinfo {author}
  {\bibfnamefont {M.}~\bibnamefont {{Garcia-Fernandez}}}, \bibinfo {author}
  {\bibfnamefont {A.}~\bibnamefont {Nag}}, \bibinfo {author} {\bibfnamefont
  {A.}~\bibnamefont {Walters}}, \bibinfo {author} {\bibfnamefont {K.-J.}\
  \bibnamefont {Zhou}}, \bibinfo {author} {\bibfnamefont {I.-M.}\ \bibnamefont
  {Pietsch}}, \bibinfo {author} {\bibfnamefont {A.}~\bibnamefont {Jesche}},
  \bibinfo {author} {\bibfnamefont {P.}~\bibnamefont {Gegenwart}}, \bibinfo
  {author} {\bibfnamefont {R.}~\bibnamefont {Coldea}}, \bibinfo {author}
  {\bibfnamefont {T.}~\bibnamefont {Schmitt}}, \ and\ \bibinfo {author}
  {\bibfnamefont {D.~F.}\ \bibnamefont {McMorrow}},\ }\href {\doibase
  10.1103/PhysRevB.100.224303} {\bibfield  {journal} {\bibinfo  {journal}
  {Phys. Rev. B}\ }\textbf {\bibinfo {volume} {100}},\ \bibinfo {pages}
  {224303} (\bibinfo {year} {2019})}\BibitemShut {NoStop}%
\bibitem [{\citenamefont {Warzanowski}\ \emph {et~al.}(2020)\citenamefont
  {Warzanowski}, \citenamefont {Borgwardt}, \citenamefont {Hopfer},
  \citenamefont {Attig}, \citenamefont {Koethe}, \citenamefont {Becker},
  \citenamefont {Tsurkan}, \citenamefont {Loidl}, \citenamefont {Hermanns},
  \citenamefont {{van Loosdrecht}},\ and\ \citenamefont
  {Gr{\"u}ninger}}]{warzanowskiMultipleSpinorbitExcitons2020b}%
  \BibitemOpen
  \bibfield  {author} {\bibinfo {author} {\bibfnamefont {P.}~\bibnamefont
  {Warzanowski}}, \bibinfo {author} {\bibfnamefont {N.}~\bibnamefont
  {Borgwardt}}, \bibinfo {author} {\bibfnamefont {K.}~\bibnamefont {Hopfer}},
  \bibinfo {author} {\bibfnamefont {J.}~\bibnamefont {Attig}}, \bibinfo
  {author} {\bibfnamefont {T.~C.}\ \bibnamefont {Koethe}}, \bibinfo {author}
  {\bibfnamefont {P.}~\bibnamefont {Becker}}, \bibinfo {author} {\bibfnamefont
  {V.}~\bibnamefont {Tsurkan}}, \bibinfo {author} {\bibfnamefont
  {A.}~\bibnamefont {Loidl}}, \bibinfo {author} {\bibfnamefont
  {M.}~\bibnamefont {Hermanns}}, \bibinfo {author} {\bibfnamefont {P.~H.~M.}\
  \bibnamefont {{van Loosdrecht}}}, \ and\ \bibinfo {author} {\bibfnamefont
  {M.}~\bibnamefont {Gr{\"u}ninger}},\ }\href {\doibase
  10.1103/PhysRevResearch.2.042007} {\bibfield  {journal} {\bibinfo  {journal}
  {Phys. Rev. Research}\ }\textbf {\bibinfo {volume} {2}},\ \bibinfo {pages}
  {042007} (\bibinfo {year} {2020})}\BibitemShut {NoStop}%
\bibitem [{\citenamefont {Suzuki}\ \emph {et~al.}(2021)\citenamefont {Suzuki},
  \citenamefont {Liu}, \citenamefont {Bertinshaw}, \citenamefont {Ueda},
  \citenamefont {Kim}, \citenamefont {Laha}, \citenamefont {Weber},
  \citenamefont {Yang}, \citenamefont {Wang}, \citenamefont {Takahashi},
  \citenamefont {F{\"u}rsich}, \citenamefont {Minola}, \citenamefont {Lotsch},
  \citenamefont {Kim}, \citenamefont {Yava{\c s}}, \citenamefont {Daghofer},
  \citenamefont {Chaloupka}, \citenamefont {Khaliullin}, \citenamefont
  {Gretarsson},\ and\ \citenamefont
  {Keimer}}]{suzukiProximateFerromagneticState2021a}%
  \BibitemOpen
  \bibfield  {author} {\bibinfo {author} {\bibfnamefont {H.}~\bibnamefont
  {Suzuki}}, \bibinfo {author} {\bibfnamefont {H.}~\bibnamefont {Liu}},
  \bibinfo {author} {\bibfnamefont {J.}~\bibnamefont {Bertinshaw}}, \bibinfo
  {author} {\bibfnamefont {K.}~\bibnamefont {Ueda}}, \bibinfo {author}
  {\bibfnamefont {H.}~\bibnamefont {Kim}}, \bibinfo {author} {\bibfnamefont
  {S.}~\bibnamefont {Laha}}, \bibinfo {author} {\bibfnamefont {D.}~\bibnamefont
  {Weber}}, \bibinfo {author} {\bibfnamefont {Z.}~\bibnamefont {Yang}},
  \bibinfo {author} {\bibfnamefont {L.}~\bibnamefont {Wang}}, \bibinfo {author}
  {\bibfnamefont {H.}~\bibnamefont {Takahashi}}, \bibinfo {author}
  {\bibfnamefont {K.}~\bibnamefont {F{\"u}rsich}}, \bibinfo {author}
  {\bibfnamefont {M.}~\bibnamefont {Minola}}, \bibinfo {author} {\bibfnamefont
  {B.~V.}\ \bibnamefont {Lotsch}}, \bibinfo {author} {\bibfnamefont {B.~J.}\
  \bibnamefont {Kim}}, \bibinfo {author} {\bibfnamefont {H.}~\bibnamefont
  {Yava{\c s}}}, \bibinfo {author} {\bibfnamefont {M.}~\bibnamefont
  {Daghofer}}, \bibinfo {author} {\bibfnamefont {J.}~\bibnamefont {Chaloupka}},
  \bibinfo {author} {\bibfnamefont {G.}~\bibnamefont {Khaliullin}}, \bibinfo
  {author} {\bibfnamefont {H.}~\bibnamefont {Gretarsson}}, \ and\ \bibinfo
  {author} {\bibfnamefont {B.}~\bibnamefont {Keimer}},\ }\href {\doibase
  10.1038/s41467-021-24722-4} {\bibfield  {journal} {\bibinfo  {journal} {Nat
  Commun}\ }\textbf {\bibinfo {volume} {12}},\ \bibinfo {pages} {4512}
  (\bibinfo {year} {2021})}\BibitemShut {NoStop}%
\bibitem [{\citenamefont {Lebert}\ \emph {et~al.}(2020)\citenamefont {Lebert},
  \citenamefont {Kim}, \citenamefont {Bisogni}, \citenamefont {Jarrige},
  \citenamefont {Barbour},\ and\ \citenamefont
  {Kim}}]{lebertResonantInelasticXray2020}%
  \BibitemOpen
  \bibfield  {author} {\bibinfo {author} {\bibfnamefont {B.~W.}\ \bibnamefont
  {Lebert}}, \bibinfo {author} {\bibfnamefont {S.}~\bibnamefont {Kim}},
  \bibinfo {author} {\bibfnamefont {V.}~\bibnamefont {Bisogni}}, \bibinfo
  {author} {\bibfnamefont {I.}~\bibnamefont {Jarrige}}, \bibinfo {author}
  {\bibfnamefont {A.~M.}\ \bibnamefont {Barbour}}, \ and\ \bibinfo {author}
  {\bibfnamefont {Y.-J.}\ \bibnamefont {Kim}},\ }\href {\doibase
  10.1088/1361-648X/ab5595} {\bibfield  {journal} {\bibinfo  {journal} {J.
  Phys.: Condens. Matter}\ }\textbf {\bibinfo {volume} {32}},\ \bibinfo {pages}
  {144001} (\bibinfo {year} {2020})}\BibitemShut {NoStop}%
\bibitem [{\citenamefont {Guizzetti}\ \emph {et~al.}(1979)\citenamefont
  {Guizzetti}, \citenamefont {Reguzzoni},\ and\ \citenamefont
  {Pollini}}]{guizzettiFundamentalOpticalProperties1979}%
  \BibitemOpen
  \bibfield  {author} {\bibinfo {author} {\bibfnamefont {G.}~\bibnamefont
  {Guizzetti}}, \bibinfo {author} {\bibfnamefont {E.}~\bibnamefont
  {Reguzzoni}}, \ and\ \bibinfo {author} {\bibfnamefont {I.}~\bibnamefont
  {Pollini}},\ }\href {\doibase 10.1016/0375-9601(79)90319-0} {\bibfield
  {journal} {\bibinfo  {journal} {Physics Letters A}\ }\textbf {\bibinfo
  {volume} {70}},\ \bibinfo {pages} {34} (\bibinfo {year} {1979})}\BibitemShut
  {NoStop}%
\bibitem [{\citenamefont {Rojas}\ and\ \citenamefont
  {Spinolo}(1983)}]{rojasHallEffectARuCl31983}%
  \BibitemOpen
  \bibfield  {author} {\bibinfo {author} {\bibfnamefont {S.}~\bibnamefont
  {Rojas}}\ and\ \bibinfo {author} {\bibfnamefont {G.}~\bibnamefont
  {Spinolo}},\ }\href {\doibase 10.1016/0038-1098(83)90738-X} {\bibfield
  {journal} {\bibinfo  {journal} {Solid State Communications}\ }\textbf
  {\bibinfo {volume} {48}},\ \bibinfo {pages} {349} (\bibinfo {year}
  {1983})}\BibitemShut {NoStop}%
\bibitem [{\citenamefont
  {Pollini}(1994)}]{polliniPhotoemissionStudyElectronic1994}%
  \BibitemOpen
  \bibfield  {author} {\bibinfo {author} {\bibfnamefont {I.}~\bibnamefont
  {Pollini}},\ }\href {\doibase 10.1103/PhysRevB.50.2095} {\bibfield  {journal}
  {\bibinfo  {journal} {Phys. Rev. B}\ }\textbf {\bibinfo {volume} {50}},\
  \bibinfo {pages} {2095} (\bibinfo {year} {1994})}\BibitemShut {NoStop}%
\bibitem [{\citenamefont {Kim}\ \emph {et~al.}(2015)\citenamefont {Kim},
  \citenamefont {V.}, \citenamefont {Catuneanu},\ and\ \citenamefont
  {Kee}}]{kimKitaevMagnetismHoneycomb2015a}%
  \BibitemOpen
  \bibfield  {author} {\bibinfo {author} {\bibfnamefont {H.-S.}\ \bibnamefont
  {Kim}}, \bibinfo {author} {\bibfnamefont {V.~S.}\ \bibnamefont {V.}},
  \bibinfo {author} {\bibfnamefont {A.}~\bibnamefont {Catuneanu}}, \ and\
  \bibinfo {author} {\bibfnamefont {H.-Y.}\ \bibnamefont {Kee}},\ }\href
  {\doibase 10.1103/PhysRevB.91.241110} {\bibfield  {journal} {\bibinfo
  {journal} {Phys. Rev. B}\ }\textbf {\bibinfo {volume} {91}},\ \bibinfo
  {pages} {241110} (\bibinfo {year} {2015})}\BibitemShut {NoStop}%
\bibitem [{\citenamefont {Sandilands}\ \emph
  {et~al.}(2016{\natexlab{a}})\citenamefont {Sandilands}, \citenamefont {Sohn},
  \citenamefont {Park}, \citenamefont {Kim}, \citenamefont {Kim}, \citenamefont
  {Sears}, \citenamefont {Kim},\ and\ \citenamefont {Noh}}]{Sandilands2016}%
  \BibitemOpen
  \bibfield  {author} {\bibinfo {author} {\bibfnamefont {L.~J.}\ \bibnamefont
  {Sandilands}}, \bibinfo {author} {\bibfnamefont {C.~H.}\ \bibnamefont
  {Sohn}}, \bibinfo {author} {\bibfnamefont {H.~J.}\ \bibnamefont {Park}},
  \bibinfo {author} {\bibfnamefont {S.~Y.}\ \bibnamefont {Kim}}, \bibinfo
  {author} {\bibfnamefont {K.~W.}\ \bibnamefont {Kim}}, \bibinfo {author}
  {\bibfnamefont {J.~A.}\ \bibnamefont {Sears}}, \bibinfo {author}
  {\bibfnamefont {Y.-J.}\ \bibnamefont {Kim}}, \ and\ \bibinfo {author}
  {\bibfnamefont {T.~W.}\ \bibnamefont {Noh}},\ }\href {\doibase
  10.1103/PhysRevB.94.195156} {\bibfield  {journal} {\bibinfo  {journal} {Phys.
  Rev. B}\ }\textbf {\bibinfo {volume} {94}},\ \bibinfo {pages} {195156}
  (\bibinfo {year} {2016}{\natexlab{a}})}\BibitemShut {NoStop}%
\bibitem [{\citenamefont {Sandilands}\ \emph
  {et~al.}(2016{\natexlab{b}})\citenamefont {Sandilands}, \citenamefont {Tian},
  \citenamefont {Reijnders}, \citenamefont {Kim}, \citenamefont {Plumb},
  \citenamefont {Kim}, \citenamefont {Kee},\ and\ \citenamefont
  {Burch}}]{sandilandsSpinorbitExcitationsElectronic2016a}%
  \BibitemOpen
  \bibfield  {author} {\bibinfo {author} {\bibfnamefont {L.~J.}\ \bibnamefont
  {Sandilands}}, \bibinfo {author} {\bibfnamefont {Y.}~\bibnamefont {Tian}},
  \bibinfo {author} {\bibfnamefont {A.~A.}\ \bibnamefont {Reijnders}}, \bibinfo
  {author} {\bibfnamefont {H.-S.}\ \bibnamefont {Kim}}, \bibinfo {author}
  {\bibfnamefont {K.~W.}\ \bibnamefont {Plumb}}, \bibinfo {author}
  {\bibfnamefont {Y.-J.}\ \bibnamefont {Kim}}, \bibinfo {author} {\bibfnamefont
  {H.-Y.}\ \bibnamefont {Kee}}, \ and\ \bibinfo {author} {\bibfnamefont
  {K.~S.}\ \bibnamefont {Burch}},\ }\href {\doibase 10.1103/PhysRevB.93.075144}
  {\bibfield  {journal} {\bibinfo  {journal} {Phys. Rev. B}\ }\textbf {\bibinfo
  {volume} {93}},\ \bibinfo {pages} {075144} (\bibinfo {year}
  {2016}{\natexlab{b}})}\BibitemShut {NoStop}%
\bibitem [{\citenamefont {Koitzsch}\ \emph {et~al.}(2016)\citenamefont
  {Koitzsch}, \citenamefont {Habenicht}, \citenamefont {M{\"u}ller},
  \citenamefont {Knupfer}, \citenamefont {B{\"u}chner}, \citenamefont
  {Kandpal}, \citenamefont {{van den Brink}}, \citenamefont {Nowak},
  \citenamefont {Isaeva},\ and\ \citenamefont
  {Doert}}]{koitzschEffDescriptionHoneycomb2016a}%
  \BibitemOpen
  \bibfield  {author} {\bibinfo {author} {\bibfnamefont {A.}~\bibnamefont
  {Koitzsch}}, \bibinfo {author} {\bibfnamefont {C.}~\bibnamefont {Habenicht}},
  \bibinfo {author} {\bibfnamefont {E.}~\bibnamefont {M{\"u}ller}}, \bibinfo
  {author} {\bibfnamefont {M.}~\bibnamefont {Knupfer}}, \bibinfo {author}
  {\bibfnamefont {B.}~\bibnamefont {B{\"u}chner}}, \bibinfo {author}
  {\bibfnamefont {H.~C.}\ \bibnamefont {Kandpal}}, \bibinfo {author}
  {\bibfnamefont {J.}~\bibnamefont {{van den Brink}}}, \bibinfo {author}
  {\bibfnamefont {D.}~\bibnamefont {Nowak}}, \bibinfo {author} {\bibfnamefont
  {A.}~\bibnamefont {Isaeva}}, \ and\ \bibinfo {author} {\bibfnamefont
  {T.}~\bibnamefont {Doert}},\ }\href {\doibase 10.1103/PhysRevLett.117.126403}
  {\bibfield  {journal} {\bibinfo  {journal} {Phys. Rev. Lett.}\ }\textbf
  {\bibinfo {volume} {117}},\ \bibinfo {pages} {126403} (\bibinfo {year}
  {2016})}\BibitemShut {NoStop}%
\bibitem [{\citenamefont {Sinn}\ \emph {et~al.}(2016)\citenamefont {Sinn},
  \citenamefont {Kim}, \citenamefont {Kim}, \citenamefont {Lee}, \citenamefont
  {Won}, \citenamefont {Oh}, \citenamefont {Han}, \citenamefont {Chang},
  \citenamefont {Hur}, \citenamefont {Sato}, \citenamefont {Park},
  \citenamefont {Kim}, \citenamefont {Kim},\ and\ \citenamefont
  {Noh}}]{Sinn2016}%
  \BibitemOpen
  \bibfield  {author} {\bibinfo {author} {\bibfnamefont {S.}~\bibnamefont
  {Sinn}}, \bibinfo {author} {\bibfnamefont {C.~H.}\ \bibnamefont {Kim}},
  \bibinfo {author} {\bibfnamefont {B.~H.}\ \bibnamefont {Kim}}, \bibinfo
  {author} {\bibfnamefont {K.~D.}\ \bibnamefont {Lee}}, \bibinfo {author}
  {\bibfnamefont {C.~J.}\ \bibnamefont {Won}}, \bibinfo {author} {\bibfnamefont
  {J.~S.}\ \bibnamefont {Oh}}, \bibinfo {author} {\bibfnamefont
  {M.}~\bibnamefont {Han}}, \bibinfo {author} {\bibfnamefont {Y.~J.}\
  \bibnamefont {Chang}}, \bibinfo {author} {\bibfnamefont {N.}~\bibnamefont
  {Hur}}, \bibinfo {author} {\bibfnamefont {H.}~\bibnamefont {Sato}}, \bibinfo
  {author} {\bibfnamefont {B.-G.}\ \bibnamefont {Park}}, \bibinfo {author}
  {\bibfnamefont {C.}~\bibnamefont {Kim}}, \bibinfo {author} {\bibfnamefont
  {H.-D.}\ \bibnamefont {Kim}}, \ and\ \bibinfo {author} {\bibfnamefont
  {T.~W.}\ \bibnamefont {Noh}},\ }\href {\doibase 10.1038/srep39544} {\bibfield
   {journal} {\bibinfo  {journal} {Sci Rep}\ }\textbf {\bibinfo {volume} {6}},\
  \bibinfo {pages} {39544} (\bibinfo {year} {2016})}\BibitemShut {NoStop}%
\bibitem [{\citenamefont {Ziatdinov}\ \emph {et~al.}(2016)\citenamefont
  {Ziatdinov}, \citenamefont {Banerjee}, \citenamefont {Maksov}, \citenamefont
  {Berlijn}, \citenamefont {Zhou}, \citenamefont {Cao}, \citenamefont {Yan},
  \citenamefont {Bridges}, \citenamefont {Mandrus}, \citenamefont {Nagler},
  \citenamefont {Baddorf},\ and\ \citenamefont
  {Kalinin}}]{ziatdinovAtomicscaleObservationStructural2016a}%
  \BibitemOpen
  \bibfield  {author} {\bibinfo {author} {\bibfnamefont {M.}~\bibnamefont
  {Ziatdinov}}, \bibinfo {author} {\bibfnamefont {A.}~\bibnamefont {Banerjee}},
  \bibinfo {author} {\bibfnamefont {A.}~\bibnamefont {Maksov}}, \bibinfo
  {author} {\bibfnamefont {T.}~\bibnamefont {Berlijn}}, \bibinfo {author}
  {\bibfnamefont {W.}~\bibnamefont {Zhou}}, \bibinfo {author} {\bibfnamefont
  {H.~B.}\ \bibnamefont {Cao}}, \bibinfo {author} {\bibfnamefont {J.-Q.}\
  \bibnamefont {Yan}}, \bibinfo {author} {\bibfnamefont {C.~A.}\ \bibnamefont
  {Bridges}}, \bibinfo {author} {\bibfnamefont {D.~G.}\ \bibnamefont
  {Mandrus}}, \bibinfo {author} {\bibfnamefont {S.~E.}\ \bibnamefont {Nagler}},
  \bibinfo {author} {\bibfnamefont {A.~P.}\ \bibnamefont {Baddorf}}, \ and\
  \bibinfo {author} {\bibfnamefont {S.~V.}\ \bibnamefont {Kalinin}},\ }\href
  {\doibase 10.1038/ncomms13774} {\bibfield  {journal} {\bibinfo  {journal}
  {Nat Commun}\ }\textbf {\bibinfo {volume} {7}},\ \bibinfo {pages} {13774}
  (\bibinfo {year} {2016})}\BibitemShut {NoStop}%
\bibitem [{\citenamefont {Zhou}\ \emph {et~al.}(2016)\citenamefont {Zhou},
  \citenamefont {Li}, \citenamefont {Waugh}, \citenamefont {Parham},
  \citenamefont {Kim}, \citenamefont {Sears}, \citenamefont {Gomes},
  \citenamefont {Kee}, \citenamefont {Kim},\ and\ \citenamefont
  {Dessau}}]{zhouAngleresolvedPhotoemissionStudy2016b}%
  \BibitemOpen
  \bibfield  {author} {\bibinfo {author} {\bibfnamefont {X.}~\bibnamefont
  {Zhou}}, \bibinfo {author} {\bibfnamefont {H.}~\bibnamefont {Li}}, \bibinfo
  {author} {\bibfnamefont {J.~A.}\ \bibnamefont {Waugh}}, \bibinfo {author}
  {\bibfnamefont {S.}~\bibnamefont {Parham}}, \bibinfo {author} {\bibfnamefont
  {H.-S.}\ \bibnamefont {Kim}}, \bibinfo {author} {\bibfnamefont {J.~A.}\
  \bibnamefont {Sears}}, \bibinfo {author} {\bibfnamefont {A.}~\bibnamefont
  {Gomes}}, \bibinfo {author} {\bibfnamefont {H.-Y.}\ \bibnamefont {Kee}},
  \bibinfo {author} {\bibfnamefont {Y.-J.}\ \bibnamefont {Kim}}, \ and\
  \bibinfo {author} {\bibfnamefont {D.~S.}\ \bibnamefont {Dessau}},\ }\href
  {\doibase 10.1103/PhysRevB.94.161106} {\bibfield  {journal} {\bibinfo
  {journal} {Phys. Rev. B}\ }\textbf {\bibinfo {volume} {94}},\ \bibinfo
  {pages} {161106} (\bibinfo {year} {2016})}\BibitemShut {NoStop}%
\bibitem [{\citenamefont {Reschke}\ \emph {et~al.}(2017)\citenamefont
  {Reschke}, \citenamefont {Mayr}, \citenamefont {Wang}, \citenamefont {Do},
  \citenamefont {Choi},\ and\ \citenamefont
  {Loidl}}]{reschkeElectronicPhononExcitations2017a}%
  \BibitemOpen
  \bibfield  {author} {\bibinfo {author} {\bibfnamefont {S.}~\bibnamefont
  {Reschke}}, \bibinfo {author} {\bibfnamefont {F.}~\bibnamefont {Mayr}},
  \bibinfo {author} {\bibfnamefont {Z.}~\bibnamefont {Wang}}, \bibinfo {author}
  {\bibfnamefont {S.-H.}\ \bibnamefont {Do}}, \bibinfo {author} {\bibfnamefont
  {K.-Y.}\ \bibnamefont {Choi}}, \ and\ \bibinfo {author} {\bibfnamefont
  {A.}~\bibnamefont {Loidl}},\ }\href {\doibase 10.1103/PhysRevB.96.165120}
  {\bibfield  {journal} {\bibinfo  {journal} {Phys. Rev. B}\ }\textbf {\bibinfo
  {volume} {96}},\ \bibinfo {pages} {165120} (\bibinfo {year}
  {2017})}\BibitemShut {NoStop}%
\bibitem [{\citenamefont {Biesner}\ \emph {et~al.}(2018)\citenamefont
  {Biesner}, \citenamefont {Biswas}, \citenamefont {Li}, \citenamefont {Saito},
  \citenamefont {Pustogow}, \citenamefont {Altmeyer}, \citenamefont {Wolter},
  \citenamefont {B{\"u}chner}, \citenamefont {Roslova}, \citenamefont {Doert},
  \citenamefont {Winter}, \citenamefont {Valent{\'i}},\ and\ \citenamefont
  {Dressel}}]{biesnerDetuningHoneycombRuCl2018}%
  \BibitemOpen
  \bibfield  {author} {\bibinfo {author} {\bibfnamefont {T.}~\bibnamefont
  {Biesner}}, \bibinfo {author} {\bibfnamefont {S.}~\bibnamefont {Biswas}},
  \bibinfo {author} {\bibfnamefont {W.}~\bibnamefont {Li}}, \bibinfo {author}
  {\bibfnamefont {Y.}~\bibnamefont {Saito}}, \bibinfo {author} {\bibfnamefont
  {A.}~\bibnamefont {Pustogow}}, \bibinfo {author} {\bibfnamefont
  {M.}~\bibnamefont {Altmeyer}}, \bibinfo {author} {\bibfnamefont {A.~U.~B.}\
  \bibnamefont {Wolter}}, \bibinfo {author} {\bibfnamefont {B.}~\bibnamefont
  {B{\"u}chner}}, \bibinfo {author} {\bibfnamefont {M.}~\bibnamefont
  {Roslova}}, \bibinfo {author} {\bibfnamefont {T.}~\bibnamefont {Doert}},
  \bibinfo {author} {\bibfnamefont {S.~M.}\ \bibnamefont {Winter}}, \bibinfo
  {author} {\bibfnamefont {R.}~\bibnamefont {Valent{\'i}}}, \ and\ \bibinfo
  {author} {\bibfnamefont {M.}~\bibnamefont {Dressel}},\ }\href {\doibase
  10.1103/PhysRevB.97.220401} {\bibfield  {journal} {\bibinfo  {journal} {Phys.
  Rev. B}\ }\textbf {\bibinfo {volume} {97}},\ \bibinfo {pages} {220401}
  (\bibinfo {year} {2018})}\BibitemShut {NoStop}%
\bibitem [{\citenamefont {Koitzsch}\ \emph {et~al.}(2020)\citenamefont
  {Koitzsch}, \citenamefont {M{\"u}ller}, \citenamefont {Knupfer},
  \citenamefont {B{\"u}chner}, \citenamefont {Nowak}, \citenamefont {Isaeva},
  \citenamefont {Doert}, \citenamefont {Gr{\"u}ninger}, \citenamefont
  {Nishimoto},\ and\ \citenamefont {{van den
  Brink}}}]{koitzschLowtemperatureEnhancementFerromagnetic2020}%
  \BibitemOpen
  \bibfield  {author} {\bibinfo {author} {\bibfnamefont {A.}~\bibnamefont
  {Koitzsch}}, \bibinfo {author} {\bibfnamefont {E.}~\bibnamefont
  {M{\"u}ller}}, \bibinfo {author} {\bibfnamefont {M.}~\bibnamefont {Knupfer}},
  \bibinfo {author} {\bibfnamefont {B.}~\bibnamefont {B{\"u}chner}}, \bibinfo
  {author} {\bibfnamefont {D.}~\bibnamefont {Nowak}}, \bibinfo {author}
  {\bibfnamefont {A.}~\bibnamefont {Isaeva}}, \bibinfo {author} {\bibfnamefont
  {T.}~\bibnamefont {Doert}}, \bibinfo {author} {\bibfnamefont
  {M.}~\bibnamefont {Gr{\"u}ninger}}, \bibinfo {author} {\bibfnamefont
  {S.}~\bibnamefont {Nishimoto}}, \ and\ \bibinfo {author} {\bibfnamefont
  {J.}~\bibnamefont {{van den Brink}}},\ }\href {\doibase
  10.1103/PhysRevMaterials.4.094408} {\bibfield  {journal} {\bibinfo  {journal}
  {Phys. Rev. Materials}\ }\textbf {\bibinfo {volume} {4}},\ \bibinfo {pages}
  {094408} (\bibinfo {year} {2020})}\BibitemShut {NoStop}%
\bibitem [{\citenamefont {Reschke}\ \emph {et~al.}(2018)\citenamefont
  {Reschke}, \citenamefont {Mayr}, \citenamefont {Widmann}, \citenamefont {{von
  Nidda}}, \citenamefont {Tsurkan}, \citenamefont {Eremin}, \citenamefont {Do},
  \citenamefont {Choi}, \citenamefont {Wang},\ and\ \citenamefont
  {Loidl}}]{reschkeSubgapOpticalResponse2018a}%
  \BibitemOpen
  \bibfield  {author} {\bibinfo {author} {\bibfnamefont {S.}~\bibnamefont
  {Reschke}}, \bibinfo {author} {\bibfnamefont {F.}~\bibnamefont {Mayr}},
  \bibinfo {author} {\bibfnamefont {S.}~\bibnamefont {Widmann}}, \bibinfo
  {author} {\bibfnamefont {H.-A.~K.}\ \bibnamefont {{von Nidda}}}, \bibinfo
  {author} {\bibfnamefont {V.}~\bibnamefont {Tsurkan}}, \bibinfo {author}
  {\bibfnamefont {M.~V.}\ \bibnamefont {Eremin}}, \bibinfo {author}
  {\bibfnamefont {S.-H.}\ \bibnamefont {Do}}, \bibinfo {author} {\bibfnamefont
  {K.-Y.}\ \bibnamefont {Choi}}, \bibinfo {author} {\bibfnamefont
  {Z.}~\bibnamefont {Wang}}, \ and\ \bibinfo {author} {\bibfnamefont
  {A.}~\bibnamefont {Loidl}},\ }\href {\doibase 10.1088/1361-648X/aae805}
  {\bibfield  {journal} {\bibinfo  {journal} {J. Phys.: Condens. Matter}\
  }\textbf {\bibinfo {volume} {30}},\ \bibinfo {pages} {475604} (\bibinfo
  {year} {2018})}\BibitemShut {NoStop}%
\bibitem [{\citenamefont {Nevola}\ \emph {et~al.}(2021)\citenamefont {Nevola},
  \citenamefont {Bataller}, \citenamefont {Kumar}, \citenamefont {Sridhar},
  \citenamefont {Frick}, \citenamefont {O'Donnell}, \citenamefont {Ade},
  \citenamefont {Maggard}, \citenamefont {Kemper}, \citenamefont {Gundogdu},\
  and\ \citenamefont {Dougherty}}]{nevolaTimescalesExcitedState2021}%
  \BibitemOpen
  \bibfield  {author} {\bibinfo {author} {\bibfnamefont {D.}~\bibnamefont
  {Nevola}}, \bibinfo {author} {\bibfnamefont {A.}~\bibnamefont {Bataller}},
  \bibinfo {author} {\bibfnamefont {A.}~\bibnamefont {Kumar}}, \bibinfo
  {author} {\bibfnamefont {S.}~\bibnamefont {Sridhar}}, \bibinfo {author}
  {\bibfnamefont {J.}~\bibnamefont {Frick}}, \bibinfo {author} {\bibfnamefont
  {S.}~\bibnamefont {O'Donnell}}, \bibinfo {author} {\bibfnamefont
  {H.}~\bibnamefont {Ade}}, \bibinfo {author} {\bibfnamefont {P.~A.}\
  \bibnamefont {Maggard}}, \bibinfo {author} {\bibfnamefont {A.~F.}\
  \bibnamefont {Kemper}}, \bibinfo {author} {\bibfnamefont {K.}~\bibnamefont
  {Gundogdu}}, \ and\ \bibinfo {author} {\bibfnamefont {D.~B.}\ \bibnamefont
  {Dougherty}},\ }\href {\doibase 10.1103/PhysRevB.103.245105} {\bibfield
  {journal} {\bibinfo  {journal} {Phys. Rev. B}\ }\textbf {\bibinfo {volume}
  {103}},\ \bibinfo {pages} {245105} (\bibinfo {year} {2021})}\BibitemShut
  {NoStop}%
\bibitem [{\citenamefont {Chaloupka}\ and\ \citenamefont
  {Khaliullin}(2016)}]{chaloupkaMagneticAnisotropyKitaev2016b}%
  \BibitemOpen
  \bibfield  {author} {\bibinfo {author} {\bibfnamefont {J.}~\bibnamefont
  {Chaloupka}}\ and\ \bibinfo {author} {\bibfnamefont {G.}~\bibnamefont
  {Khaliullin}},\ }\href {\doibase 10.1103/PhysRevB.94.064435} {\bibfield
  {journal} {\bibinfo  {journal} {Phys. Rev. B}\ }\textbf {\bibinfo {volume}
  {94}},\ \bibinfo {pages} {064435} (\bibinfo {year} {2016})}\BibitemShut
  {NoStop}%
\bibitem [{\citenamefont {de~la Torre}\ \emph {et~al.}(2021)\citenamefont
  {de~la Torre}, \citenamefont {Zager}, \citenamefont {Bahrami}, \citenamefont
  {DiScala}, \citenamefont {Chamorro}, \citenamefont {Upton}, \citenamefont
  {Fabbris}, \citenamefont {Haskel}, \citenamefont {Casa}, \citenamefont
  {McQueen}, \citenamefont {Tafti},\ and\ \citenamefont
  {Plumb}}]{delatorre2021}%
  \BibitemOpen
  \bibfield  {author} {\bibinfo {author} {\bibfnamefont {A.}~\bibnamefont
  {de~la Torre}}, \bibinfo {author} {\bibfnamefont {B.}~\bibnamefont {Zager}},
  \bibinfo {author} {\bibfnamefont {F.}~\bibnamefont {Bahrami}}, \bibinfo
  {author} {\bibfnamefont {M.}~\bibnamefont {DiScala}}, \bibinfo {author}
  {\bibfnamefont {J.~R.}\ \bibnamefont {Chamorro}}, \bibinfo {author}
  {\bibfnamefont {M.~H.}\ \bibnamefont {Upton}}, \bibinfo {author}
  {\bibfnamefont {G.}~\bibnamefont {Fabbris}}, \bibinfo {author} {\bibfnamefont
  {D.}~\bibnamefont {Haskel}}, \bibinfo {author} {\bibfnamefont
  {D.}~\bibnamefont {Casa}}, \bibinfo {author} {\bibfnamefont {T.~M.}\
  \bibnamefont {McQueen}}, \bibinfo {author} {\bibfnamefont {F.}~\bibnamefont
  {Tafti}}, \ and\ \bibinfo {author} {\bibfnamefont {K.~W.}\ \bibnamefont
  {Plumb}},\ }\href {\doibase 10.1103/PhysRevB.104.L100416} {\bibfield
  {journal} {\bibinfo  {journal} {Phys. Rev. B}\ }\textbf {\bibinfo {volume}
  {104}},\ \bibinfo {pages} {L100416} (\bibinfo {year} {2021})}\BibitemShut
  {NoStop}%
\bibitem [{\citenamefont {Gretarsson}\ \emph
  {et~al.}(2013{\natexlab{b}})\citenamefont {Gretarsson}, \citenamefont
  {Clancy}, \citenamefont {Singh}, \citenamefont {Gegenwart}, \citenamefont
  {Hill}, \citenamefont {Kim}, \citenamefont {Upton}, \citenamefont {Said},
  \citenamefont {Casa}, \citenamefont {Gog},\ and\ \citenamefont
  {Kim}}]{gretarssonMagneticExcitationSpectrum2013a}%
  \BibitemOpen
  \bibfield  {author} {\bibinfo {author} {\bibfnamefont {H.}~\bibnamefont
  {Gretarsson}}, \bibinfo {author} {\bibfnamefont {J.~P.}\ \bibnamefont
  {Clancy}}, \bibinfo {author} {\bibfnamefont {Y.}~\bibnamefont {Singh}},
  \bibinfo {author} {\bibfnamefont {P.}~\bibnamefont {Gegenwart}}, \bibinfo
  {author} {\bibfnamefont {J.~P.}\ \bibnamefont {Hill}}, \bibinfo {author}
  {\bibfnamefont {J.}~\bibnamefont {Kim}}, \bibinfo {author} {\bibfnamefont
  {M.~H.}\ \bibnamefont {Upton}}, \bibinfo {author} {\bibfnamefont {A.~H.}\
  \bibnamefont {Said}}, \bibinfo {author} {\bibfnamefont {D.}~\bibnamefont
  {Casa}}, \bibinfo {author} {\bibfnamefont {T.}~\bibnamefont {Gog}}, \ and\
  \bibinfo {author} {\bibfnamefont {Y.-J.}\ \bibnamefont {Kim}},\ }\href
  {\doibase 10.1103/PhysRevB.87.220407} {\bibfield  {journal} {\bibinfo
  {journal} {Phys. Rev. B}\ }\textbf {\bibinfo {volume} {87}},\ \bibinfo
  {pages} {220407} (\bibinfo {year} {2013}{\natexlab{b}})}\BibitemShut
  {NoStop}%
\bibitem [{\citenamefont {Kim}\ \emph {et~al.}(2020)\citenamefont {Kim},
  \citenamefont {Chaloupka}, \citenamefont {Singh}, \citenamefont {Kim},
  \citenamefont {Kim}, \citenamefont {Casa}, \citenamefont {Said},
  \citenamefont {Huang},\ and\ \citenamefont
  {Gog}}]{kimDynamicSpinCorrelations2020a}%
  \BibitemOpen
  \bibfield  {author} {\bibinfo {author} {\bibfnamefont {J.}~\bibnamefont
  {Kim}}, \bibinfo {author} {\bibfnamefont {J.}~\bibnamefont {Chaloupka}},
  \bibinfo {author} {\bibfnamefont {Y.}~\bibnamefont {Singh}}, \bibinfo
  {author} {\bibfnamefont {J.~W.}\ \bibnamefont {Kim}}, \bibinfo {author}
  {\bibfnamefont {B.~J.}\ \bibnamefont {Kim}}, \bibinfo {author} {\bibfnamefont
  {D.}~\bibnamefont {Casa}}, \bibinfo {author} {\bibfnamefont {A.}~\bibnamefont
  {Said}}, \bibinfo {author} {\bibfnamefont {X.}~\bibnamefont {Huang}}, \ and\
  \bibinfo {author} {\bibfnamefont {T.}~\bibnamefont {Gog}},\ }\href {\doibase
  10.1103/PhysRevX.10.021034} {\bibfield  {journal} {\bibinfo  {journal} {Phys.
  Rev. X}\ }\textbf {\bibinfo {volume} {10}},\ \bibinfo {pages} {021034}
  (\bibinfo {year} {2020})}\BibitemShut {NoStop}%
\end{thebibliography}%

\end{document}